# Neural Network Kinetics for Exploring Diffusion Multiplicity and Chemical Ordering in Compositionally Complex Materials

**Bin Xing[1,2], Timothy J. Rupert[1,2], Xiaoqing Pan[1,2], Penghui Cao[1,3,2*]**

[1]Center for Complex and Active Materials, University of California, Irvine, Irvine, California 92697, United States

[2]Department of Material Science and Engineering, University of California Irvine, Irvine, California 92697, United States

[3]Department of Mechanical and Aerospace Engineering, University of California, Irvine, Irvine, CA 92697, United States

* Email: caoph@uci.edu

## Abstract

**Diffusion involving atom transport from one location to another governs many important processes and behaviors such as precipitation and phase nucleation. Local chemical complexity in compositionally complex materials poses challenges for modeling atomic diffusion and the resulting formation of chemically ordered structures. Here, we introduce a neural network kinetics (NNK) scheme that predicts and simulates diffusion-induced chemical and structural evolution in complex concentrated chemical environments. The framework is grounded on efficient on-lattice structure and chemistry representation combined with artificial neural networks, enabling precise prediction of all path-dependent migration barriers and individual atom jumps. To demonstrate the method, we study the temperature-dependent local chemical ordering in a refractory Nb-Mo-Ta alloy and reveal a critical temperature at which the B2 order reaches a maximum. The atomic jump randomness map exhibits the highest diffusion heterogeneity (multiplicity) in the vicinity of this characteristic temperature, which is closely related to chemical ordering and B2 structure formation. The scalable NNK framework provides a promising new avenue to exploring diffusion-related properties in the vast compositional space within which extraordinary properties are hidden.**

## Introduction

Diffusion in materials dictates the kinetics of precipitation[1], new phase formation[2] and microstructure evolution[3], and strongly influences mechanical and physical properties[4]. For example, altering nanoprecipitate size and dispersion by thermal processing enables substantial increases in strength and good ductility in multicomponent alloys[5,6]. Essentially rooted in diffusion kinetics, predicting how fast local composition and microstructure evolve is a fundamental goal of material science. In metals and alloys, diffusion processes are connected with vacancies, point defects that mediate atom jumps in the crystal lattice. Molecular dynamics (MD)[7] modeling based on force fields or density functional theory, which probe the atomic mechanisms of diffusion at a nanosecond timescale, are often not able to access slow diffusion kinetics-induced microstructure change. To circumvent this time limitation inherent in MD, the kinetic Monte Carlo method (kMC) is primarily adopted to model diffusion-mediated structure evolution, for instance, the early stage of precipitation in dilute alloys[8,9]. In the kMC simulations, the crucial parameter (vacancy migration energy) is generally parameterized from continuum models such as cluster expansion[10] and Ising model[11], owing to the high computational cost in transition state search. The rise of compositionally complex alloys (CCAs), commonly known as high-entropy alloys, brings many intriguing kinetics behaviors, ranging from chemical short-range ordering[12], precipitation[6], segregation[13], and radiation defect annihilation[14], which have yet to be fundamentally interpreted and ultimately predicted. The chemical complexity of such events, however, poses a new challenge for modeling diffusion-mediated processes due to local chemical fluctuations leading to diverse activation barriers (i.e., a wide spectrum)[15].

The emergence of machine learning methods has demonstrated the potential for addressing computationally complex problems in materials science that involve nonlinear interactions and massive combinatorial space[16]. One of the most promising examples is machine-learned interatomic potentials that map a three-dimensional (3D) atomic configuration to its conformational energy with a high accuracy at a substantially reduced computational cost[17]. The key step in machine learning in molecular science is converting atomistic structure into numerical values (descriptive parameters–descriptor[18]) to represent the individual local chemical and structural environments. Two successful atomic environment descriptors are atom-centered symmetry function[19] and smooth overlap of atomic position[20]. The dimension of these local structure descriptors (consideration of all neighboring atoms within a cutoff distance) increases quadratically with the number of constituent elements[21], which escalates the number of parameters and training time for the application of machine learning to chemically complex CCAs. To address this issue, active efforts have been taken to compress chemical information and reduce the size of representation of local atomic environment[22–24]. Using the structure descriptor, atomic site related scalar values, such as segregation energy[25] and atomic propensity to rearrange[26], have been predicted through machine learning models. Concerning vacancy diffusion in compositionally complex alloys, a critical parameter of interest is diffusion energy barrier $\Delta E$, i.e., the energy difference between transition state and the initial energy minimum (Figure S1). Due to atomic-scale composition fluctuation and the existence of multiple diffusion directions in CCAs, it necessitates a machine learning model to precisely predict vectoral property, specifically, diffusion path-dependent barriers. Another complexity, needing to be addressed in modeling diffusion and new phase formation in CCAs, lies in the extensive compositional space and the development of local chemical order, both of which profoundly impact on diffusion barriers and kinetics.

In this study, we introduce a neural network kinetics (NNK) scheme for predicting atomic diffusion and its resulting microstructure evolution in compositionally complex materials. Grounded on an efficient on-lattice atomic representation that converts individual atoms to neurons while preserving the atomic structure, the NNK precisely describes atom (interneuron) interactions through a neural network model and predicts neuron kinetics evolution, embodying physical atom diffusion and microstructure evolution. With only one-time conversion of atomic configuration to neuron map, vacancy diffusions and chemical evolution are simulated by swapping neurons, rending high efficiency and scalability. Using refractory NbMoTa as a model system, we explore chemical ordering and B2 phase formation mediated by diffusion kinetics and reveal the anomalous diffusion (diffusion multiplicity) that is inherent in CCAs.

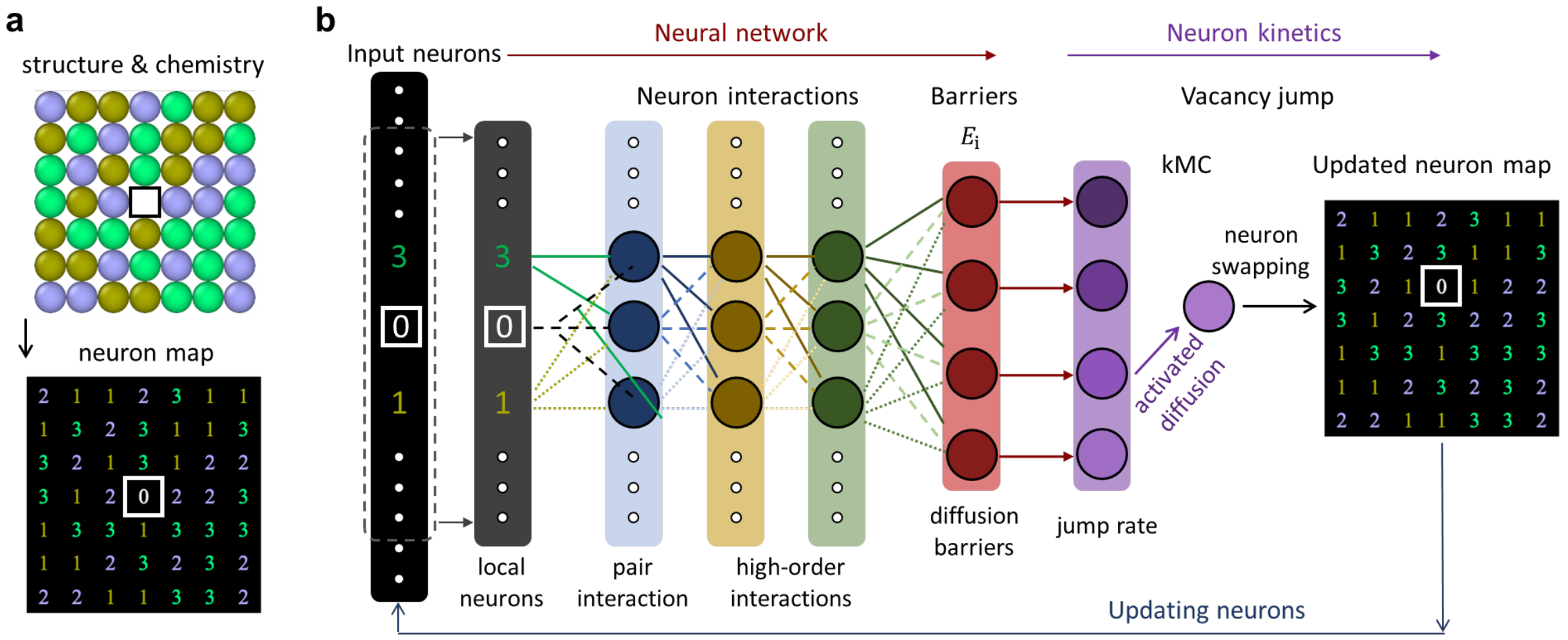

**Figure 1. Schematic illustration of neural network kinetics (NNK) framework. a**, The on-lattice structure and chemistry representation. A vacancy and its local atomic environment are encoded into a digital matrix (neuron map). **b**, NNK framework consists of a neural network that outputs vacancy migration barriers, and a neuron kinetics module that implements neuron jump (diffusion jump) based on kinetic Monte Carlo (kMC). See methods section for details on neuron kinetics. Vacancy jumps and chemical evolution are efficiently modeled by swapping of neurons and neuron map evolution.

## Results

**Neural network kinetics scheme**. Figure 1a shows the on-lattice structure and chemistry representation, where the initial atomic configuration with a vacancy is encoded into a digital matrix, or neuron map. The digits (1, 2, and 3) represent the corresponding atom types, and 0 denotes the vacancy (refer to Figure S2 for conversion and visualization of 3D crystals). This digital matrix capturing structure and composition features offers several advantages important as a descriptor[18]. The map dimension $O(N)$ scales linearly with the number of atoms $N$, which has the lowest dimension possible as descriptor. Importantly, the determination of the descriptive map is simple and involves no intensive calculation or painstaking parameter tuning. Essential for diffusion, the representation is rotational non-invariant and enables prediction of diffusion path-dependent activation barriers (vector quantities). These vectorized digits are then passed to the NNK model and serve as input neurons.

Figure 1b depicts the schematic of the NNK which consists of an artificial neural network and a neuron kinetics module. The introduced neural network (with more than two hidden layers) is designed to learn the nonlinear interactions between input neurons (i.e., atoms and vacancy), and to predict the diffusion energy barriers. Notably, the network only uses the vacancy and its neighboring neurons as inputs, resulting in a low and constant computational cost (independent of system size) without sacrificing accuracy (see Supplementary section 3 for details). With the available barriers associated with each individual diffusion path, the neuron kinetics module adopts the kinetic Monte Carlo method to carry out diffusion kinetics evolution (see Methods). There are two features rendering the NNK a high computational efficiency and scalability with system size. First, the descriptor map is calculated only once for the initial atomic configuration, because atomic diffusion and local chemical evolution are operated on the representing neuron map. Second, since atomic diffusion depends solely on the local chemical environment, the NNK trained on small configurations can be directly applied to large systems for diffusion modeling. Therefore, with only one-time conversion of atomic configuration to neuron map, vacancy jumps and chemical evolution can be simulated by swapping two digits of neural map. In this way, millions of vacancy jumps can be modeled efficiently, with each jump iteration involving the action of just two neurons (Figure 1b).

**Predicting a path-dependent diffusion barrier spectrum in multidimensional composition space**. Diffusion in crystals occurs through elementary atomic jumps between a vacancy and its neighboring lattice sites (vacancy mechanism[4,27]). In body-centered cubic (bcc) CCAs, a vacancy is associated with eight different jump directions, and the variation in the jumping atoms and surrounding chemical environment can result in eight distinct migration barriers[15,28]. By utilizing the rotational non-invariant lattice representation, it is possible to predict the jump path-dependent barriers (a vector quantity) from a single chemical configuration. Specifically, by aligning each diffusion path to a constant reference orientation through rotation and/or mirroring operations unique, neuron map and digital vector, $D_i$, can be generated for each individual diffusion path $i$, without breaking the structural symmetry, as demonstrated in Figure 2a. The Table S1 and Figure S3 summarizes the operations aligning the diffusion direction of interest with this reference, preserving structural symmetry.

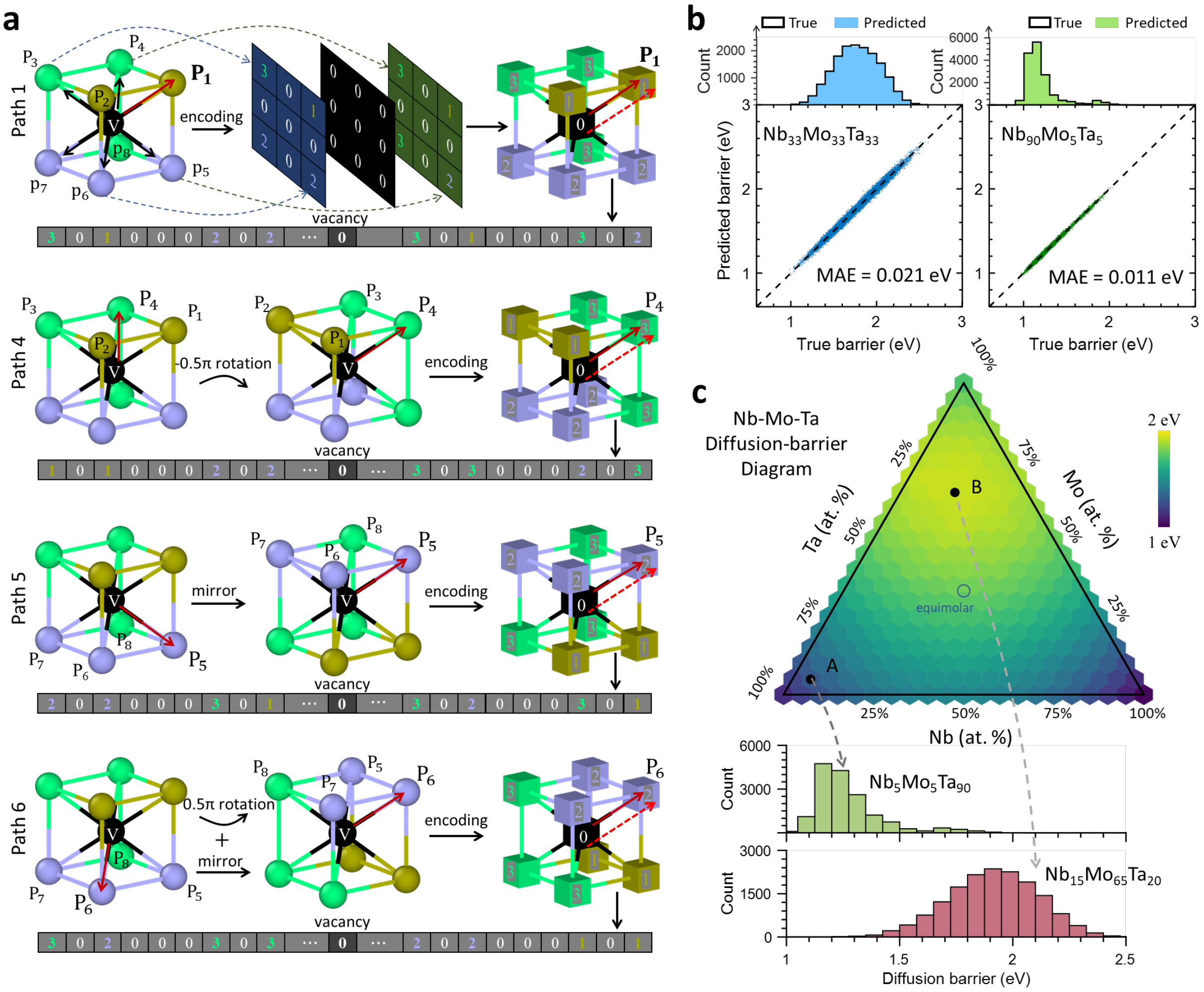

**Figure 2. Predicting diffusion barrier spectra in the entire composition space of Nb-Mo-Ta.** **a**, Creation of unique neuron maps and feature vectors for each individual diffusion path P, which enables the prediction of eight path-dependent barriers from a vacancy. The symbol V represents the vacancy. **b**, Performance of neural network in predicting diffusion barrier spectrum in concentrated, $Nb_{33}Mo_{33}Ta_{33}$, and dilute, $Nb_{90}Mo_5Ta_5$, solutions. **c**, Diffusion barrier diagram generated by the neural network. The non-equimolar $Nb_{10}Mo_{70}Ta_{20}$ alloy exhibits the highest barrier in the Nb-Mo-Ta system.

The neural network takes in $D_i$, which carries local atomic environment encompassing the vacancy, as input. The data (atomic digits) then flow through hidden layers to the output layer, which predicts the associated diffusion activation barrier, $E_i$. The first hidden layer in neural network characterizes the linear contribution of the input neurons (atoms and vacancy) to the migration barrier, while the following hidden layers capture the nonlinear and high-order interactions that impact vacancy jump. With just four hidden layers and 112 neighboring atoms (up to the 8th nearest neighbor shell) of the vacancy, the neural network achieves a high level of accuracy in predicting the path-dependent diffusion barrier (Supplementary section 4 and Figures S12-14 for the testing of different neural network structures). Figure 2b presents the evaluation of machine learning model performance for two different concentrations (one concentrated and one dilute), where the predicted energy barrier value is compared with the ground truth (see Methods). The predicted and true values exhibit the same spectrum of barriers, and the mean absolute error (MAE) is less than 1.2% of the average true migration barrier for the two alloys, concentrated solution $Nb_{33}Mo_{33}Ta_{33}$ and dilute solution $Nb_{90}Mo_5Ta_5$ (see Figure S4 for new compositions and different system sizes).

After training on only tens of compositions (Supplementary section 5 and Figures S15-16), the neural network remarkably harnesses the complete composition space of the ternary Nb-Mo-Ta system, building the relationship between composition and diffusion barrier spectrum. Figure 2c shows the diffusion barrier diagram generated by the neural network, from which the alloy ($Nb_{10}Mo_{70}Ta_{20}$) having the highest mean barrier is quickly identified. While research efforts have been primarily focused on equimolar or near-equimolar compositions, our results indicate an abnormal behavior can originate from non-equimolar concentrations hidden in the vast composition space. The neural network, which accurately predicts diffusion barriers for new and unseen compositions, implies that it fully deciphers the complex local chemistry variation and links it with diffusion property.

**Diffusion kinetics-induced local chemical order.** Originating from the attractive and repulsive interactions among the constituent elements of CCAs, atomic diffusion leads to the emergence of local chemical order on a short- to medium-range scale. To uncover diffusion-mediated chemical ordering and its dependence on annealing temperature, we employ the NNK model to simulate the equimolar NbMoTa system, using a model which contains 1,024 lattice sites, at temperatures ranging from 100 to 3000 K. With the ability to resolve individual atomic jump and the low computational cost, 20 million diffusion jumps are carried out for each temperature.

Figure 3a shows the change of the local chemical order $\delta_{ij}$ as a function of temperature. Here the non-proportional order parameter, $\delta_{ij}$, quantifies the chemical order between a pair of atom types $i$ and $j$ in the first nearest neighbor shell (see Methods). A positive $\delta_{ij}$ indicates a higher number of pairs compared to a random solid solution, suggesting that element $i$ prefers to bond with element $j$ (favored pairing), while a negative value suggests an unfavored pairing. At a high temperature (3000 K), the system ultimately approaches the random solid solution, as reflected by the small value of $\delta_{ij}$. As the temperature decreases, the magnitudes of $\delta_{ij}$ for Mo-Ta, Ta-Ta, Mo-Mo pairs increase monotonically until they reach a turning point (around 800 K), beyond which the trend reverses. The chemical order falls rapidly as the temperature is further lowered and, at 400 K, it nearly vanishes. It is noted that the system experienced an identical number of 20 million jumps at all temperatures. These results suggest the existence of a critical temperature at which the diffusion-favored ordering reaches a maximum (Regime I in Figure 3a). Below the critical value (Regime II), diffusion jumps barely develop and enhance chemical order.

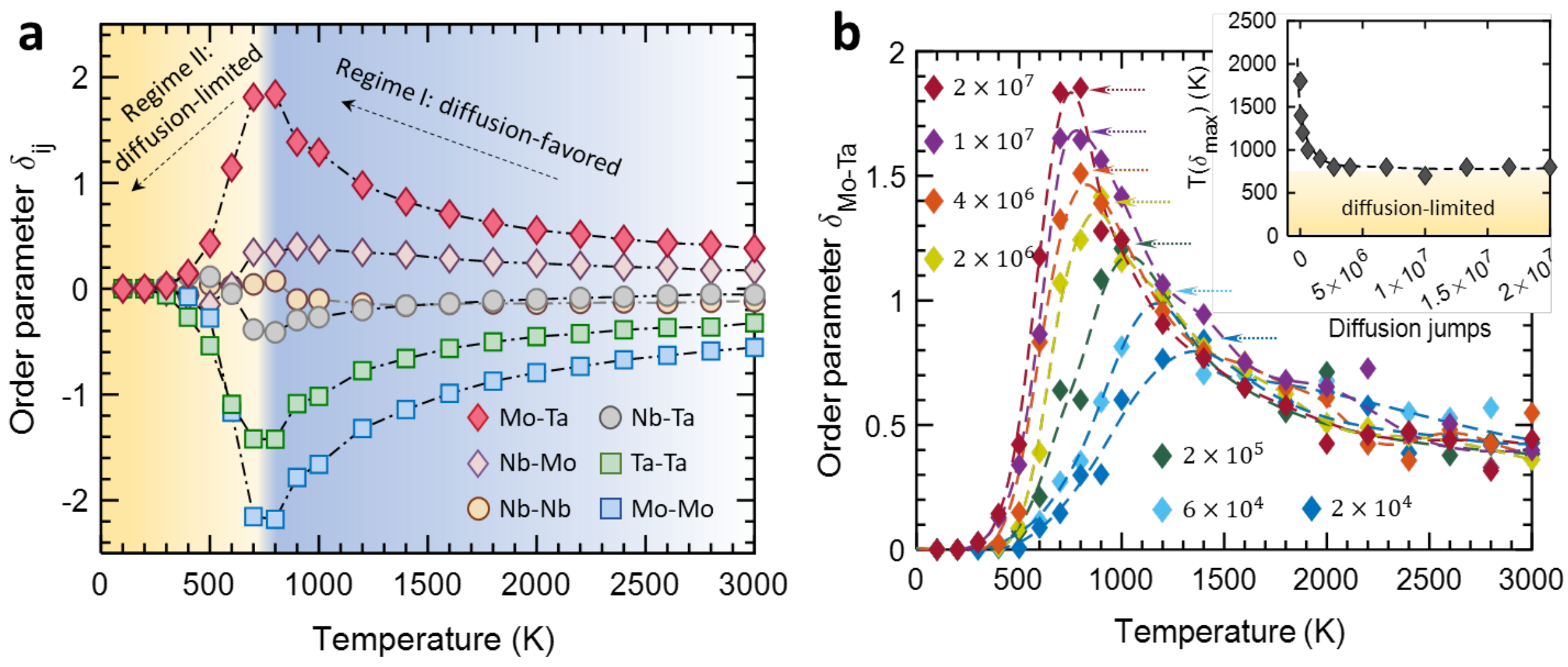

**Figure 3. Diffusion kinetics-mediated local chemical order in the equimolar NbMoTa alloy. a**, Variation of chemical order $\delta_{\mathrm{ij}}$ obtained at different annealing temperatures displays a critical temperature that divides the map into two characteristic regimes, denoted as diffusion-favored (I) and diffusion-limited (II). **b**, Development of Mo-Ta order, $\delta_{\mathrm{Mo-Ta}}$, as a function of diffusion jumps from $2 \times 10^4$ to $2 \times 10^7$. The inset shows that the jump number dependence of peak temperature suggest the critical temperature ~800 K below which the chemical ordering is suppressed.

To better understand this critical temperature and how the number of diffusion jumps affects it, we present the $\delta_{\mathrm{Mo-Ta}}$ order parameter values obtained from a wide range of jumps, from $2 \times 10^4$ to $2 \times 10^7$, in Figure 3b. As the number of jumps increases, the characteristic temperature $T(\delta_{max})$ corresponding to the maximum order gradually shifts to lower values. The inset of Figure 3b illustrates the variation of $T(\delta_{max})$ with diffusion jumps, again unveiling this critical temperature below which diffusion-mediated ordering is substantially limited.

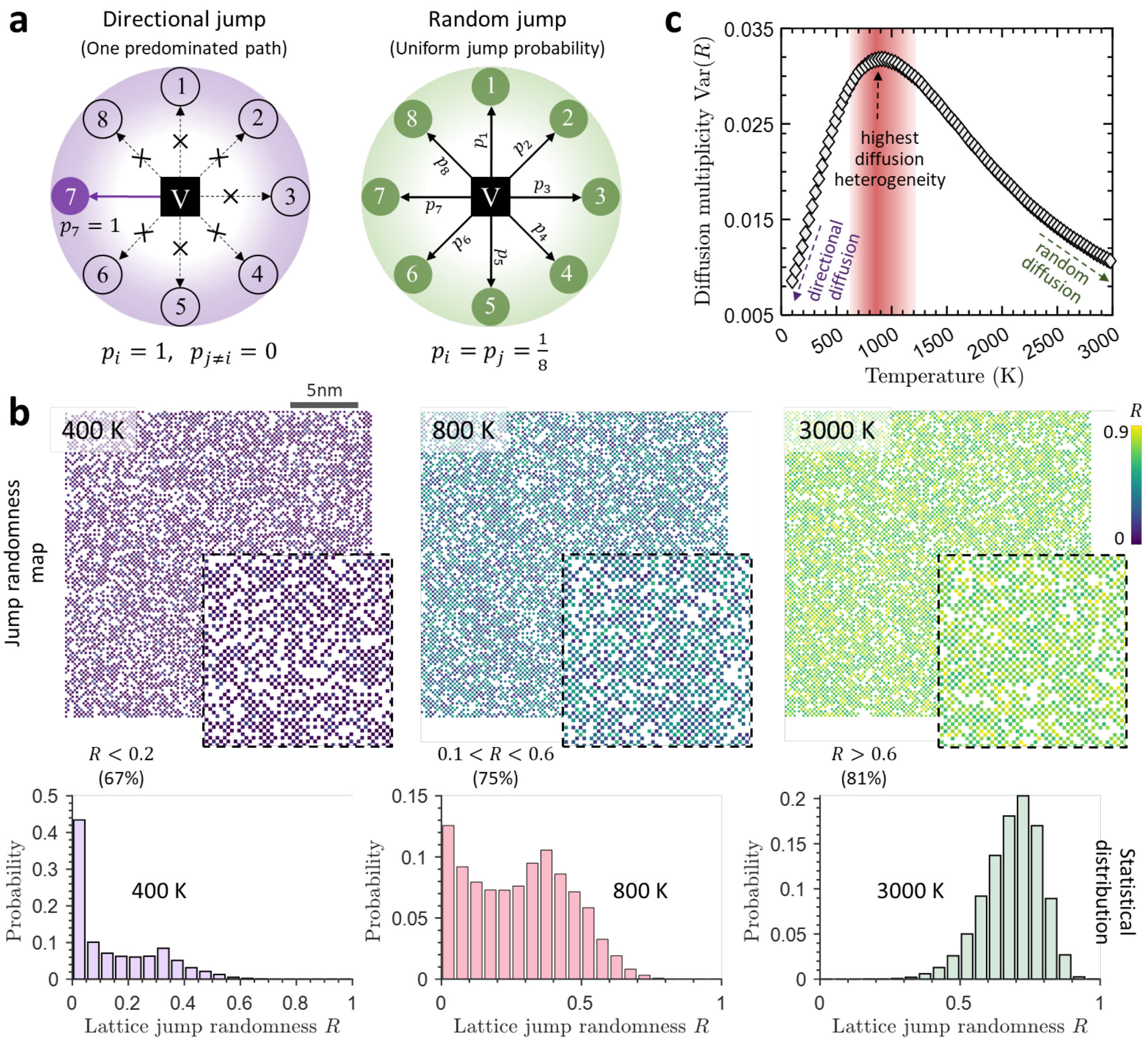

**Figure 4. Jump randomness and diffusion multiplicity of an equimolar NbMoTa alloy. a**, Schematics of two limiting lattice jump modes. One of the eight paths is predominated in directional jump (jump randomness $R = 0$), while all eight paths have the same hopping probability $p$ in random jump ($R = 1$). **b**, Spatial and statistical distributions of lattice jump randomness, $R$, at three representative temperatures. At 3000 K the distribution of $R$ ($R_{\text{peak}} = 0.7$) indicates highly random diffusion, while at 400 K the lattice jumps transform to directional diffusion mode ($R_{\text{peak}} = 0.0$). Lattice jumps at 800 K exhibit highly heterogeneous diffusion modes, shown by the broad distribution of $R$. **c**, Diffusion multiplicity $Var(R)$ as a function of temperature reveals a critical temperature (~850 K) at which diffusion is more heterogeneous (widest distribution of $R$). Moving to the two ends, diffusion approaches simple random or directions modes at ultimate high and low temperatures, respectively.

**Jump randomness and diffusion multiplicity in CCAs.** In monoatomic crystals, the diffusion of vacancy can be described as purely random, with each possible jump path having an equal probability of occurrence. However, in CCAs, local variations in chemical composition give rise to distinct and path-dependent energy barriers, resulting in a multivariate distribution of jump probabilities. For example, in bcc CCAs, the jump probability for each of the eight possible paths associated with a vacancy site can be expressed as $p_i = exp(-E_i/k_\mathrm{B}T)/\sum_{j=1}^{8} exp\left(E_j/k_\mathrm{B}T\right)$, where $E_i$ is the energy barrier of path $i$, $k_\mathrm{B}$ is Boltzmann constant, and $T$ is temperature. This can lead to various diffusion modes, as illustrated in Figure 4a, with the two limiting cases being pure random jump (where all jump paths have the same probability of occurrence) and non-random, directional lattice jump (where one path predominates). To quantify the degree of lattice jump randomness, we define an order parameter $R = 1 - \sigma(\mathbf{p})/\max(\sigma)$, where $\sigma(\mathbf{p})$ is the standard deviation of jump probability, $\mathbf{p}$, and $\max(\sigma)$ is the maximum standard deviation occurring in directional jump. Note the parameter, $R$, ranging from 0 to 1, quantifies the degree of jump randomness, with $R$ = 1 and $R$ = 0 representing the limiting cases of random diffusion and directional diffusion, respectively.

Figure 4b shows spatial and statistical distributions of lattice jump randomness $R$ at three representative temperatures. The spatial maps display color-coded lattices based on their respective $R$ values. At a high temperature of 3000 K, the thermal energy ($k_B T \gg E_i$) smears out the energy barrier difference between paths, leading to a peak $R$ value of 0.7, indicating highly random jumps. It is tempting to speculate that random atomic diffusion is insufficient to build and develop B2 ordered phase, which apparently corresponds to the low order observed at high temperatures (Figure 3a). At a low temperature of 400 K, the lattice jumps transform into directional diffusion, as demonstrated by the $R$ distribution having a peak value of 0. This implies that only one of the eight diffusion pathways is active at each lattice site. Presumably, this one-dimensional directional diffusion predominating at low temperatures (< 400 K) limits and suppresses the nucleation and growth of three-dimensional B2 structure. Intriguingly, at an intermediate temperature (~800 K), the lattice jump randomness $R$ exhibits a broad distribution, spanning from 0.0 to 0.7, indicating highly heterogeneous diffusion modes.

To assess the system-level diffusion multiplicity (heterogeneity) and its temperature dependence, we calculate the variance of diffusion randomness $\mathrm{Var}(R)$ across temperatures ranging from 100 - 3000 K, as illustrated in Figure 4c. When close to the high or low-temperature ends, there is a rapid change in $\mathrm{Var}(R)$, implying that diffusion approaches a random or directional mode. The temperature variation of $\mathrm{Var}(R)$ reveals a peak value of diffusion multiplicity at around 850 K. Random and directional-type lattice jumps are spatially interspersed throughout the entire system, as shown in the spatial map of Figure 4b. The observation of the highest diffusion multiplicity (Figure 4c) and maximum B2 order (Figure 3a) occurring in the same intermediate temperature range suggests a strong correlation between diffusion heterogeneity and the formation of local chemical order.

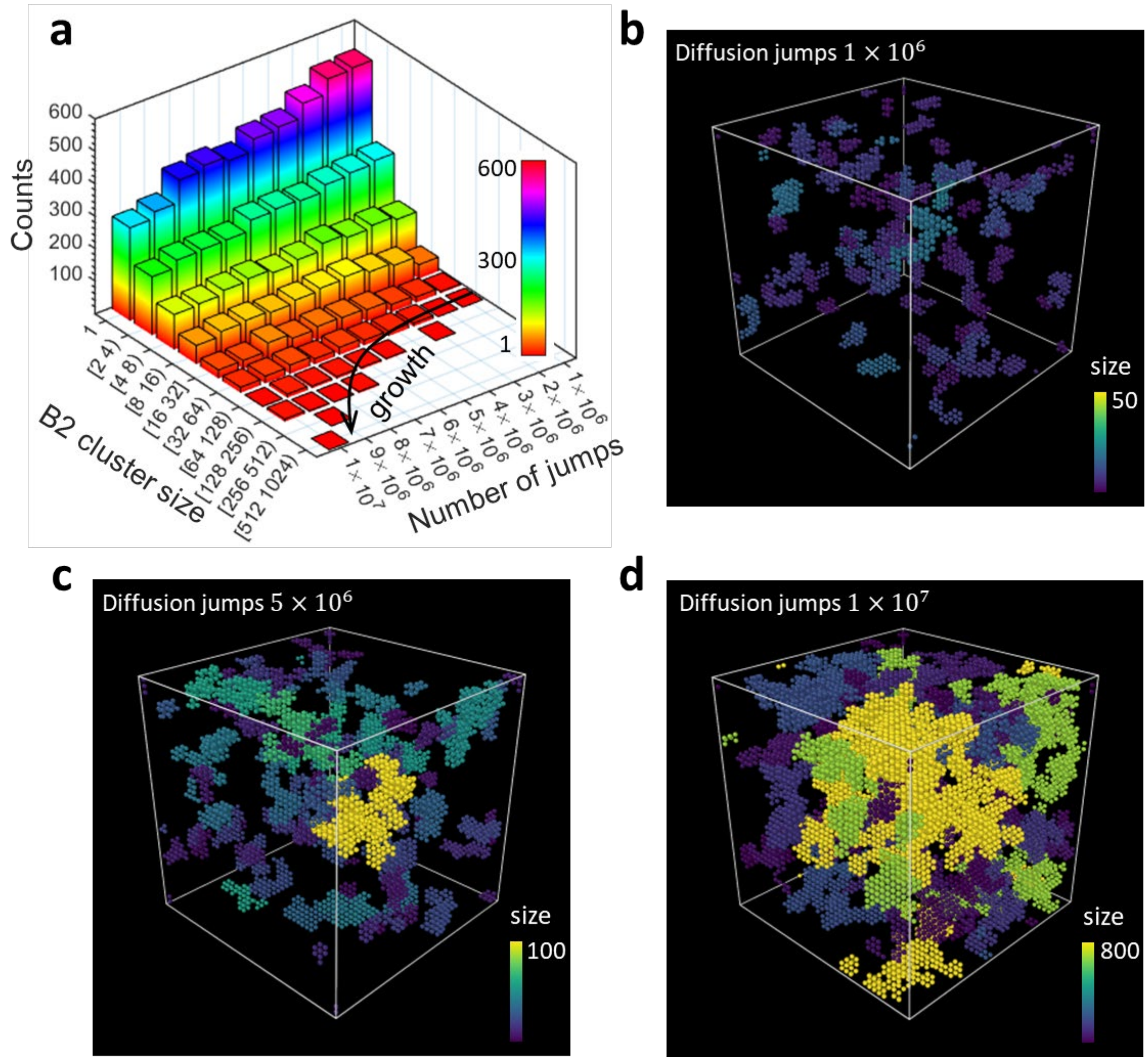

**Figure 5. B2 structure nucleation and growth kinetics during annealing in NbMoTa. a**, B2 cluster size evolution with the number of diffusion jumps. **b-d**, Spatial distributions of growing B2 cluster at $1 \times 10^6$, $5 \times 10^6$, and $1 \times 10^7$ diffusion jumps. Clusters are color coded by their size.

**B2 structure nucleation and growth kinetics.** Determining the formation kinetics of chemically ordered structure in a complex solid solution has been a challenge due to the local chemical fluctuations and huge amounts of diffusion barriers. The NNK framework efficiently and precisely predicting diffusion barrier at any chemical environment is intended to address this issue. To demonstrate the efficacy of the model, we perform aging simulations of NbMoTa consisting of 128,000 atoms. Figure 5a-c shows the spatial-temporal nucleation and evolution of B2 structure induced by diffusion. With $1 \times 10^6$ diffusion jumps, considerable amount of B2 clusters emerge in the system (Figure 5b), most of which are small clusters (size < 8 atoms). As the number of diffusion jumps further increases ($1 \times 10^7$), large clusters begin to appear and continue to grow, accompanied by annihilation and reduction of small ones (Figure 5a). The decrease in spatially isolated small clusters are a result of their attachment or adsorption by nearby growing large ones. Apart from small clusters, another essential kinetic process underlying growth is large cluster interaction and coalescence. When two spreading clusters come near to each other, they merge into a large one mediated by diffusion (Figure S8). Figure 5d reveals the spatial distribution of formed B2 clusters colored by their size in the aged material. In contrast to the precipitation of ordered nanoparticles in dilute solutions, the more heterogenous growth of chemically ordered structure signifies the substantial role of diffusion multiplicity in governing the complex chemical ordering in concentrated solutions.

## Discussion

Diffusion kinetics in the emergent compositionally complex materials[29,30] (often called high-entropy alloys and high entropy oxides) raise many intriguing rate-controlling phenomena and properties, such as chemical short-range order[12], chemically ordered nanoparticle formation[31], decomposition[32], superionic conductivity[33], extraordinary radiation tolerance[14,34], to new a few. These behaviors are controlled by the underlying atomic diffusion, which occurs in a chemical environment with a high degree of local composition fluctuations. Uncovering the kinetic processes and predicting structure evolution in these materials requires novel computational techniques that can disentangle their chemical complexity and connect it with individual atomic jumps. The NNK scheme introduced here aims to tackle the kinetic behaviors arising from diffusion processes, with a particular focus on this novel class of materials. Underpinned by an interpretable chemistry and structure representation (neuron map), the neural network precisely predicts the diffusion-path dependent energy barriers governing individual atomic jumps. The atomic diffusion and structure variations are effectively modeled on the neuron map through neuron digit exchange (Figure 1b). This framework possesses three key advantages that give both high computational efficiency and accuracy in modeling diffusion and new phase formation. Firstly, the interpretable on-lattice representation, which converts chemistry and structure to physically equivalent neuron maps, yields an ultra-small feature size, critical for machine learning models. Secondly, the determination of neuron map (descriptor) is a one-time and simple process, as it can be updated to fully replicate atomic diffusion jumps and structure evolution. Importantly, the rotational non-invariance of the neuron map enables the prediction of vector values from a single neuron map (vacancy configuration). Thirdly, the NNK trained by small models can be applied directly to investigate the kinetic behavior of large systems without sacrificing accuracy. This size scalability is demonstrated, for instance, by accurate barrier predictions (see Figure S4, S20) and ordered phase growth in large NbMoTa systems (Figure 5).

The cluster expansion (CE)[35,36] method has long been used to study thermodynamic properties of multicomponent systems, such as vacancy formation energy[37]. For diffusion kinetics, the pivotal factor is determining the diffusion barriers, requiring calculation of transition states (saddle points). While the CE has been commonly employed to predict the energies of local minimum states[8,9], presenting the transition state using CE and predicting the associated energy barrier remains a challenging task[38] (see SI section 6). Particularly, the increase in chemical complexity makes the design of clusters for even local minimum configurations a time-consuming process. To tackle this challenge, an approach involving parametrizing the reaction coordinate and minimum energy path has been proposed[39], however, leading to a low prediction accuracy. Another machine learning model promising to atomistic modeling is graph neural network (GNN), which has shown great success in developing universal machine learning interatomic potentials[40,41]. Regarding vacancy diffusion in CCAs, GNN theoretically has the potential to predict vector properties using rotation-covariance features. However, modeling vacancy jump and chemical evolution using graph network entails node swapping and updating edge properties. Each node swap (representing vacancy jump) can potentially affect neighboring nodes and their connected edge features, necessitating their updates. This requires altering a significant portion of the network, encompassing the 8th nearest neighbors of vacancy node, after each vacancy jump. In contrast, the introduced NNK scheme, with neural map representation, simplifies the process by only requiring the update of two neurons for each diffusion jump. This simplicity allows the mirroring of vacancy jumps through the swapping of neurons (digits). With just one-time conversion of the atomic configuration to a neuron map, vacancy diffusion and chemical evolution are efficiently simulated by swapping digits (the vacancy neuron and one of its nearest neighboring neurons) according to precise diffusion barriers and system temperature. In this way, tens of millions of vacancy jumps are modeled efficiently, with each jump iteration involving the action of just two neurons.

Stemming from attractive/repulsive interactions between solutes, atomic diffusion inevitably leads to nucleation of chemically ordered structure in CCAs during annealing. Using the NNK and bcc NbMoTa as model system, we uncover the existence of a critical temperature, at which the B2 order reaches its maximum value. This temperature dependence of chemical order is closely related to the underlying lattice jump randomness, as shown by the randomness maps (Figure 4). At high temperatures close to melting point, diffusion jumps ultimately approach a purely random process, corresponding to a low propensity for order formation. At low temperatures, lattice diffusion becomes dominated by the lowest barrier path, manifesting as directional jumping and restricting the nucleation of chemically ordered structure. At the critical temperature in the intermediate range, random-like and directional-type lattice jumps spread the entire system, exhibiting the highest diffusion heterogeneity (multiplicity, Figure 4c). By tracking individual B2 clusters during annealing, it is found that their nucleation and growth are intermittent and non-uniform, accompanied by the reduction and annihilation of small clusters (Figure 5 and Supplementary Video 1). This salient feature in kinetics growth of B2 structure is not captured by fictitious thermodynamics-based modeling using random atom type swap (see Methods and Figure S9), which shows a more uniform growth (Figure S10). These results highlight the complex and multitudinous kinetic pathways in CCAs towards stable states, where many processes like ordered structure nucleation, annihilation, growth, and rearrangement are interplayed and coordinated.

The neural network trained on dozens of compositions demonstrates high performance for unseen compositions, unveiling the entire ternary space of Nb-Mo-Ta (Figure 2c). With the design space for composition being practically limitless, the compositionally complex material formed by mixing multiple element opens a new frontier waiting to be explored. Traditional structure-property calculations relying on density functional theory and molecular dynamics work well for small datasets but fall short in harnessing the vast composition space. Recent advances in the rapidly growing field of machine learning creates a fertile ground for computational material science[42,43], having led to the discovery of alloys with optimal properties[44]. By directly connecting the multidimensional composition with diffusion barrier spectra, the NNK illuminates a bright path to explore the vast compositional space of CCAs, where hidden extraordinary kinetic properties lie.

# Methods

## Material system and diffusion barrier calculation

We focus on the emergent refractory CCA, Nb-Mo-Ta, as the study system to demonstrate the neural network kinetics (NNK) scheme. When generating diffusion datasets for training the neural networks, we use relatively small atomic models which has $10 \times 10 \times 10$ unit cells (containing 2000 atoms). The climbing image nudged elastic band (CI-NEB)[45] method is adopted to compute the vacancy diffusion energy barriers in the Nb-Mo-Ta system using a state-of-the-art machine learning potential[46]. For one initial configuration of a vacancy, the eight final configurations are prepared by swapping the vacancy with its first nearest neighbor atoms. By labeling each diffusion path, the path-dependent diffusion energy barriers are therefore generated. Before CI-NEB calculation, both initial and final configurations are optimized to their local energy minimum states. The CI-NEB inter-replica spring constant is set to be 5 eV/Å$^2$, and the energy tolerance and force tolerance are 0 eV and 0.01 eV/Å, respectively. The choice of parameters that optimize convergence of the calculations result in essentially the same energy barrier using smaller tolerance and large spring constant.

## Structure representation and neural networks

The on-lattice representation coverts the atomic structure into a digit matrix, which will be deciphered by neural networks. The conversion is done through a voxel grid that separates the 3D material model into uniform cubes. Each grid acquires a digit value (voxel) according to its enclosed atom type or vacancy. For bcc structure, the largest grid we can use, which can fully distinct all lattices and yield the smallest voxel grid dimensionality, is $a/2$, where $a$ is the lattice constant of the crystal (see Supplementary section 2).

The neural network, taking the representative structure and chemistry digits (neurons) as input, process them through the hidden layers, outputting the energy barriers. The connections between neurons in the hidden layers imitate the physical interactions between atoms and atom-vacancy. Representing the interaction strength (contribution to the migration barrier), the weights associated with the connections are adjusted during training. To understand the influence of network architecture on prediction performance, we train a series of neural networks with varying number of layers and number of neurons in each layer (Supplementary section 4). As the number of neurons in each hidden layer increases from 16, 32, 64, to 256, the testing MAE rapidly decreases, followed by convergence at 128 that is enough to explicitly describe all the local neighbors of a vacancy (Figure S14). By testing the different number of layers, the final network structure with 4 hidden layers and 128 neurons in each layer was selected for simulating the diffusion in the equimolar NbMoTa alloy, owing to its robustness in concentrated solid solutions. In addition, we separately train a convolutional neural network (CNN) to compare with the simple neural network. The CNN comprises four convolutional layers that compress the 3D neuroma map to $1 \times 128$ dimension for barrier prediction. The architecture of CNN is depicted in Figure S17 and described in Supplementary section 4. Likely resulting from adaptive learning spatial hierarchies of features from input 3D atomic structure, CNN exhibits slightly enhanced predictive performance (Figure S20).

The training data are generated from 46 different compositions which uniformly sample the Nb-Mo-Ta diagram (Figure S18 and Table S3). In Supplementary section 5, we carefully study and discuss the number of compositions required to train a highly accurate network for predicting the complete ternary space. Each composition model contains 2000 atoms, giving rise to 16,000 diffusion barriers. The total 736,000 data points are split into training dataset (95% total data) and validation dataset (5%). All the compositions and their data points are summarized in Table S3. After validation, the neural network is tested for barrier prediction in unseen compositions (which are not used for training or validation) and in atomic configuration with different sizes. For example, Figure S4 shows the testing results for the new compositions, $Nb_{10}Mo_{10}Ta_{80}$, $Nb_{20}Mo_{60}Ta_{20}$, $Nb_{40}Mo_{30}Ta_{30}$, and the average MAE is around 0.018 eV. Notably the neural network preserves the consistent high accuracy for different sized systems containing 512, 2000, and 6750 atoms, indicating scalability.

**Neuron kinetics**

The neuron map enables efficient modeling of vacancy kinetics through the exchange of neurons, referred to as neuron kinetics. By converting the atomic configuration into a neuron map just once, the neural network simulates vacancy jumps simply by swapping two neurons within the map (vacancy and one of its nearest neighboring neurons). This streamlined process allows to efficiently model tens of millions of vacancy jumps. Importantly, it is worth noting that each jump iteration involves the exchange of only two neurons, as depicted in Figure 1b.

Vacancy jump is carried on the neuron map based on the kinetic Monte Carlo (kMC) algorithm. Diffusion occurs through vacancy (vacancy neuron) jump to its nearest neighboring sites, and each site has a jump rate defined by $k_i = k_0 \exp(-E_i/k_B T)$, where $E_i$ is the energy barrier along jump path $i$, $k_B$ is Boltzmann constant, $T$ is temperature, and $k_0$ is an attempt frequency. The vacancy diffusion barriers associated with the eight jump paths are obtained from the neural network. The total jump rate for the current vacancy configuration is $R = \sum_{i=1}^{8} k_i$, i.e., the sum of all individual elementary rate. To simulate kinetic evolution, we first draw a uniform random number $u \in (0,1]$ and select a diffusion path, $p$, which satisfies the condition[47], $\sum_{i=1}^{p-1} k_i/R \le u \le \sum_{i=1}^{p} k_i/R$. The vacancy jump along path $p$ are then executed by exchanging the vacancy with the selected neighboring neuron (neuron digit swapping), resulting in an updated neuron map for the next iteration.

**Static Monte Carlo and molecular dynamics simulation**

We perform static Monte Carlo (MC) simulations coupled with molecular dynamics to reveal the chemical order determined by enthalpy (mainly thermodynamics). In each MC trial, a pair of atoms is randomly selected for type swap. The acceptance probability is according to the $\exp(-\Delta H/k_B T)$ in Metropolis algorithm[48]. The term $\Delta H$ is the enthalpy change after swap, therefore, the chemical evolution and ordering is predominately contorted by enthalpy. The MC swaps are followed by MD equilibration. For the systems consisting 1024 atoms, we perform 18,000 swap attempts (each atom on average subjected to 18 swaps) and 600 ps MD equilibrium. Figure S9 shows the local order as a function of MC step for temperatures from 100 - 3000 K. To study B2 cluster growth, we perform the MC and MD simulation in a large model (128,000 atoms). There are totally 135,000 swaps coupled with 150 ps MD equilibrium. Unlike diffusion-mediated B2 cluster growth, the clusters grow in a uniform and homogeneous manner (Figure S10).

**Local chemical order parameter**

To quantify the degree of chemical order, we use the non-proportional parameter[49] $\delta_{ij} = N_{ij} - N_{0,ij}$, where $N_{ij}$ denotes the actual number of pairs between atoms $i$ and $j$ in the first nearest neighboring shell, and $N_{0,ij}$ represents the average number of pairs in random solutions. A positive $\delta_{ij}$ means a favored and increased number of $i$-j pairs, indicating element $i$ tends to bond with element $j$. A negative $\delta_{ij}$ indicates unfavored pair, meaning $i$ and $j$ repel each other. Random solid solution has $\delta_{ij} = 0$.

**B2 cluster analysis**

Mo and Ta tend to attract each other and form the B2 structure. The B2 unit cell has the simple bcc structure and comprises two species, Ta and Mo, orderly located in the cube corners or center.

The unit cell can have either Ta or Mo-centered pattern. Because of the high concentration of Nb in the equimolar NbMoTa alloy, we characterize a unit as B2 when 3/4 of the Ta nearest neighbors are Mo, or 3/4 of the Mo nearest neighbors are Ta. To analysis the B2 cluster, the identified individual B2 units are gathered into individual group according to distance criterion. Two B2 units can have volume-, face-, edge-, and point-sharing at distance $\sqrt{3}a/2$, $a$, $\sqrt{2}a$, $\sqrt{3}$a (i.e., 5th shell), respectively, where $a$ is lattice constant (illustrated in Figure S7). Choosing the cut-off distance as half of the 5th shell and 6th shell, the spatial distribution and size of all B2 clusters can be successfully characterized. During the kinetic annealing, clusters can be reduced or annihilated, which causes clusters appearance or disappearance from time to time. The fluctuation hinders visualization and analysis of stable B2 cluster evolution. To address this issue, we search and identify the persist clusters that exist all the time during annealing. Focusing on the persist cluster provides a clear evolution of cluster growth (Figure 5).

## Data availability

The diffusion data in this study have been deposited in the Zenodo under accession code https://doi.org/10.5281/zenodo.7714650

## Code availability

All source codes of NNK are available at GitHub repository https://github.com/UCICaoLab/NKK[50].

## Author contributions

P.C. conceived the research idea, wrote the manuscript, and generated the figures with inputs from B.X.. B.X. developed the model, implemented the code, and performed simulation and modeling. T.J.R and X.P. reviewed and edited the manuscript. All authors contributed to data analysis and project discussion.

## Competing Interests

The author declares no competing interests.

# Supplementary Information

Neural Network Kinetics for Exploring Diffusion Multiplicity and Chemical Ordering in Compositionally Complex Materials

## Table of Contents

## 1. Supplementary Figures

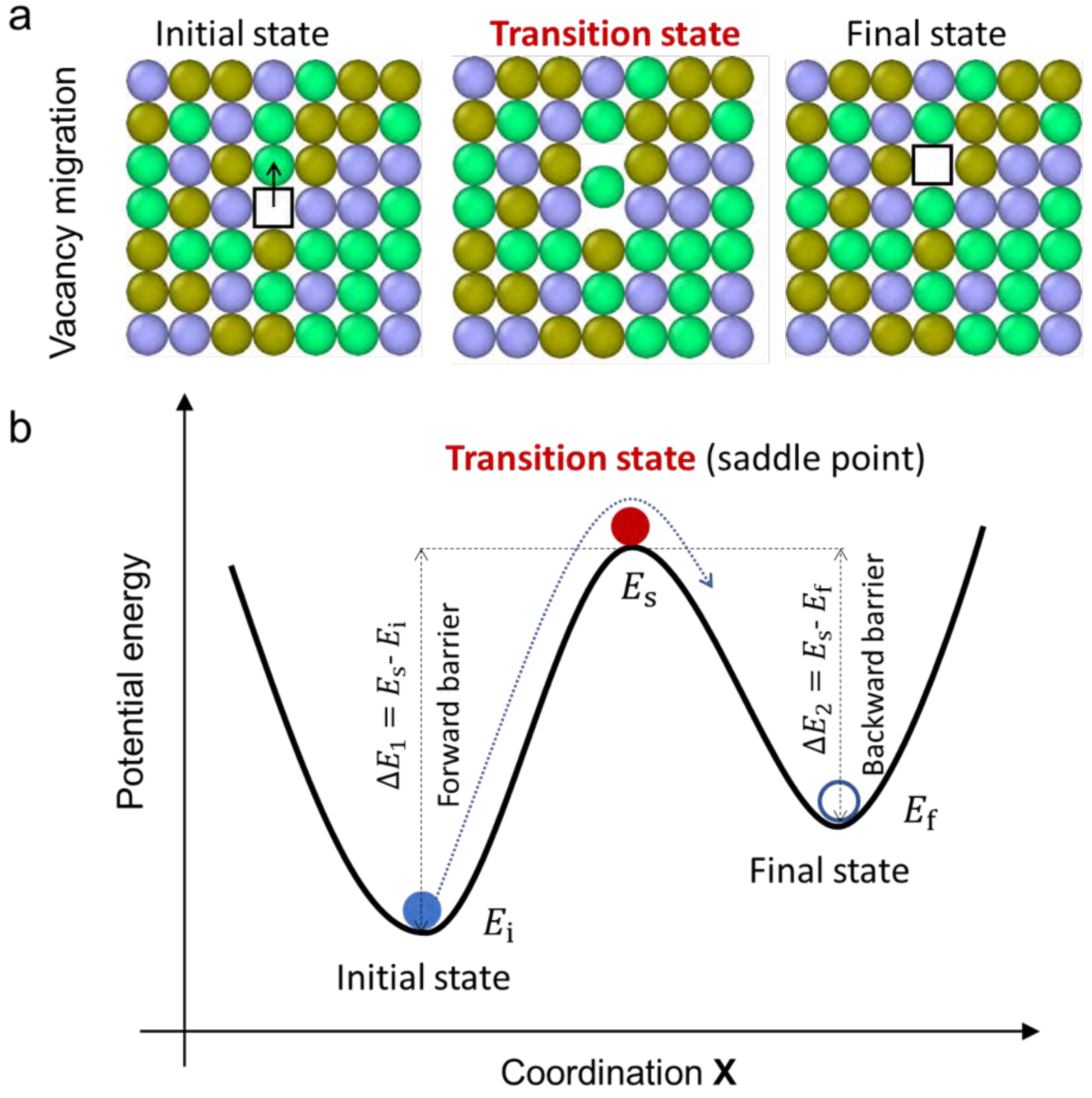

**Figure S1. Schematic illustration of vacancy diffusion and the corresponding diffusion energy landscape.** (a) Vacancy diffusion states from an initial state, through saddle point, and leads to the final state. (b) The energy barrier $\Delta E$, i.e., the energy difference between transition state and the initial energy minimum, is the governing value for diffusion. The key task is to accurately and efficiently predict these barriers in compositionally complex materials.

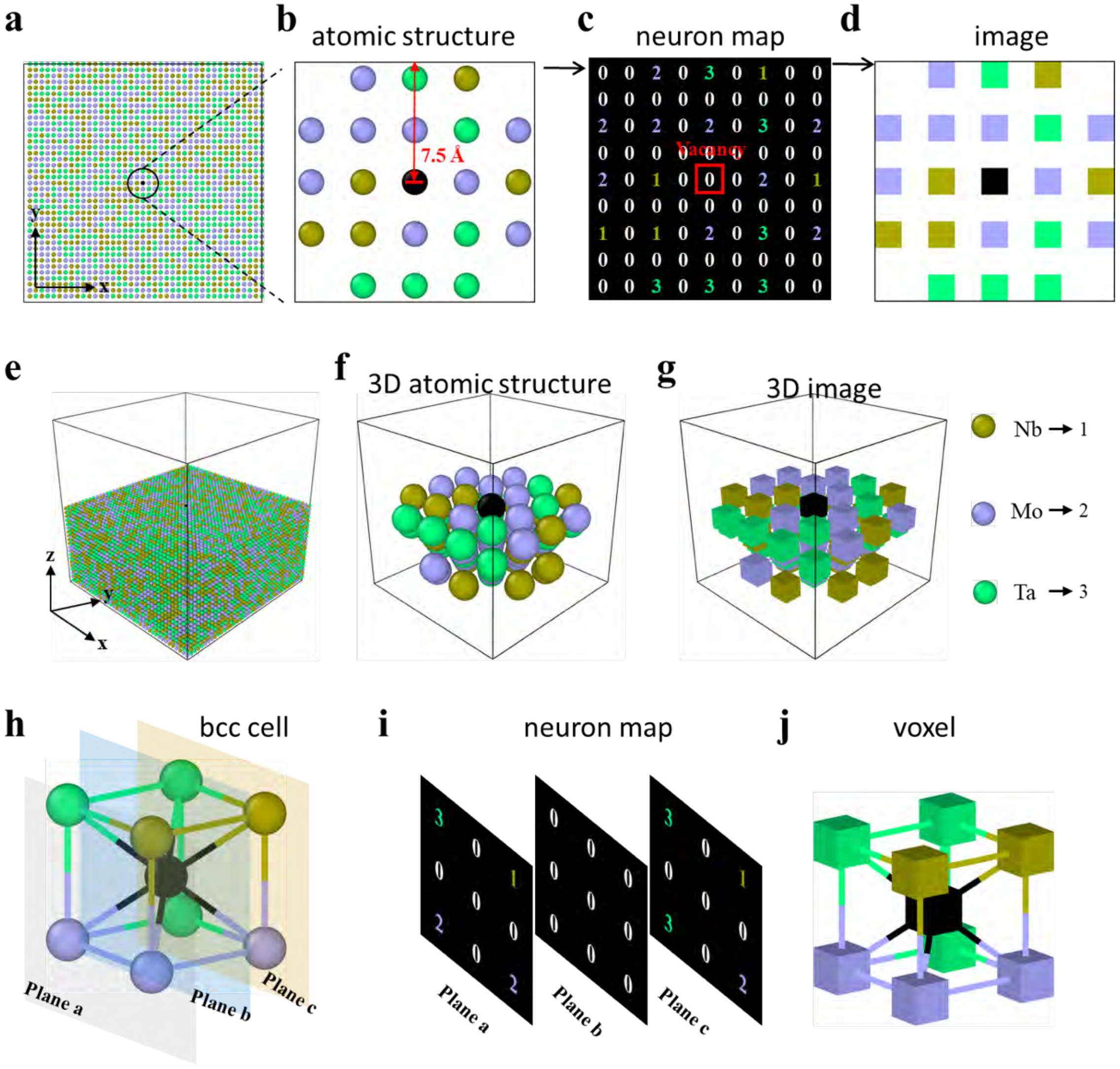

**Figure S2. On-lattice presentation of local atomic environments in equimolar NbMoTa alloy.** (a) Atom plane containing a vacancy (color-coded by black). (b) Enlarged view of the region within the circle region in (a) with cutoff distance 7.5 angstroms. (c-d) Digit matrix (neuron map) converted from atomic structure. (e-g) 3D illustration of atomic configuration within/below the vacancy-containing layer. (h) Vacancy and its first nearest neighboring atoms, and (i-j) the corresponding neuron map. The nearest neighbors are determined based on Euclidean distance between vacancy and atoms.

**Table S1. Operations of aligning eight diffusion pathways with the reference direction.**

| Path | Rotation | Mirror | Path | Rotation | Mirror |
|---|---|---|---|---|---|
| V-1 | 0 | No | V-5 | 0 | Yes |
| V-2 | 0.5π | No | V-6 | 0.5π | Yes |
| V-3 | π | No | V-7 | π | Yes |
| V-4 | -0.5π | No | V-8 | -0.5π | Yes |

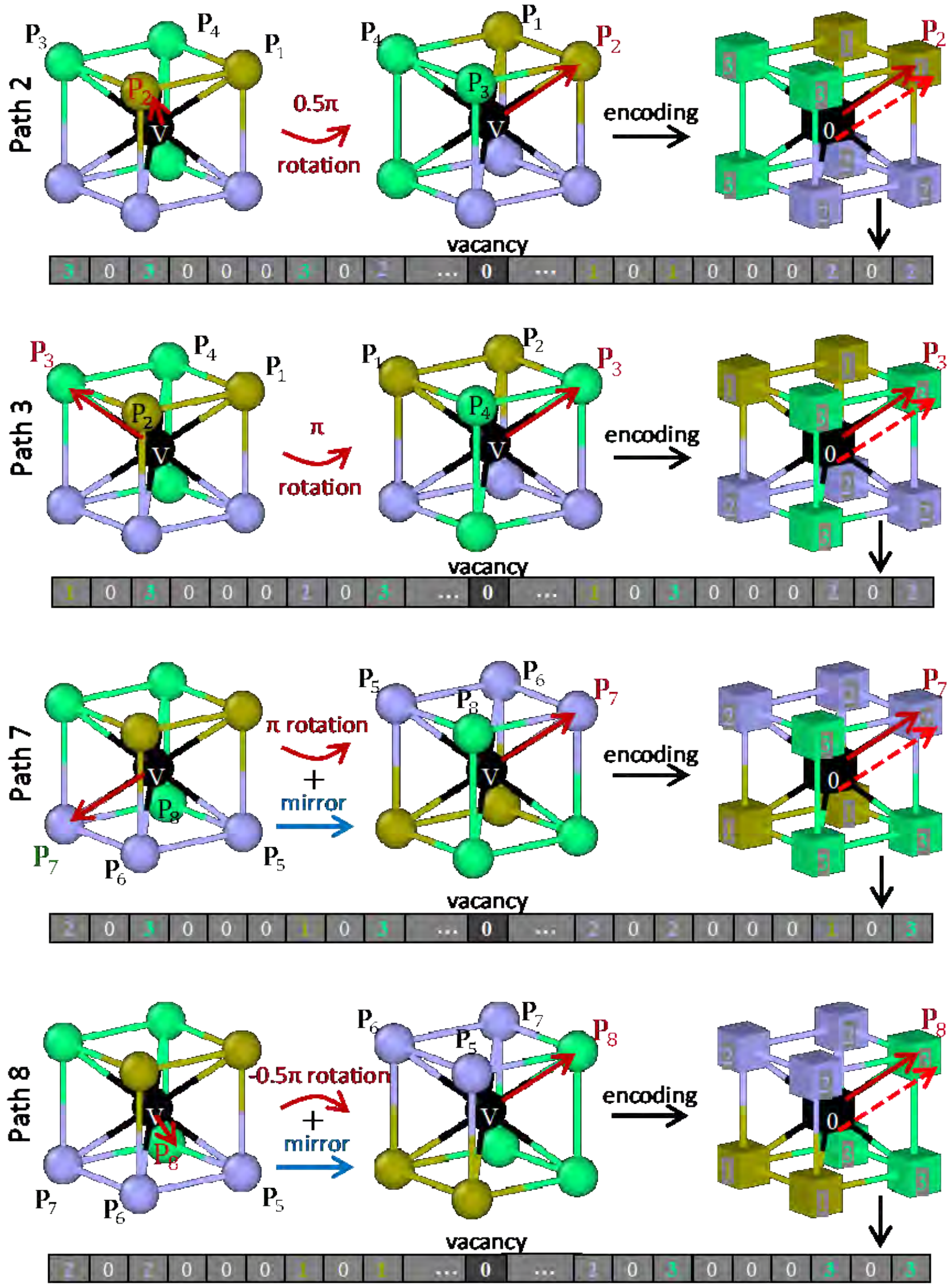

**Figure S3.** Aligning diffusion pathways 2, 3, 7 and 8 with the reference direction.

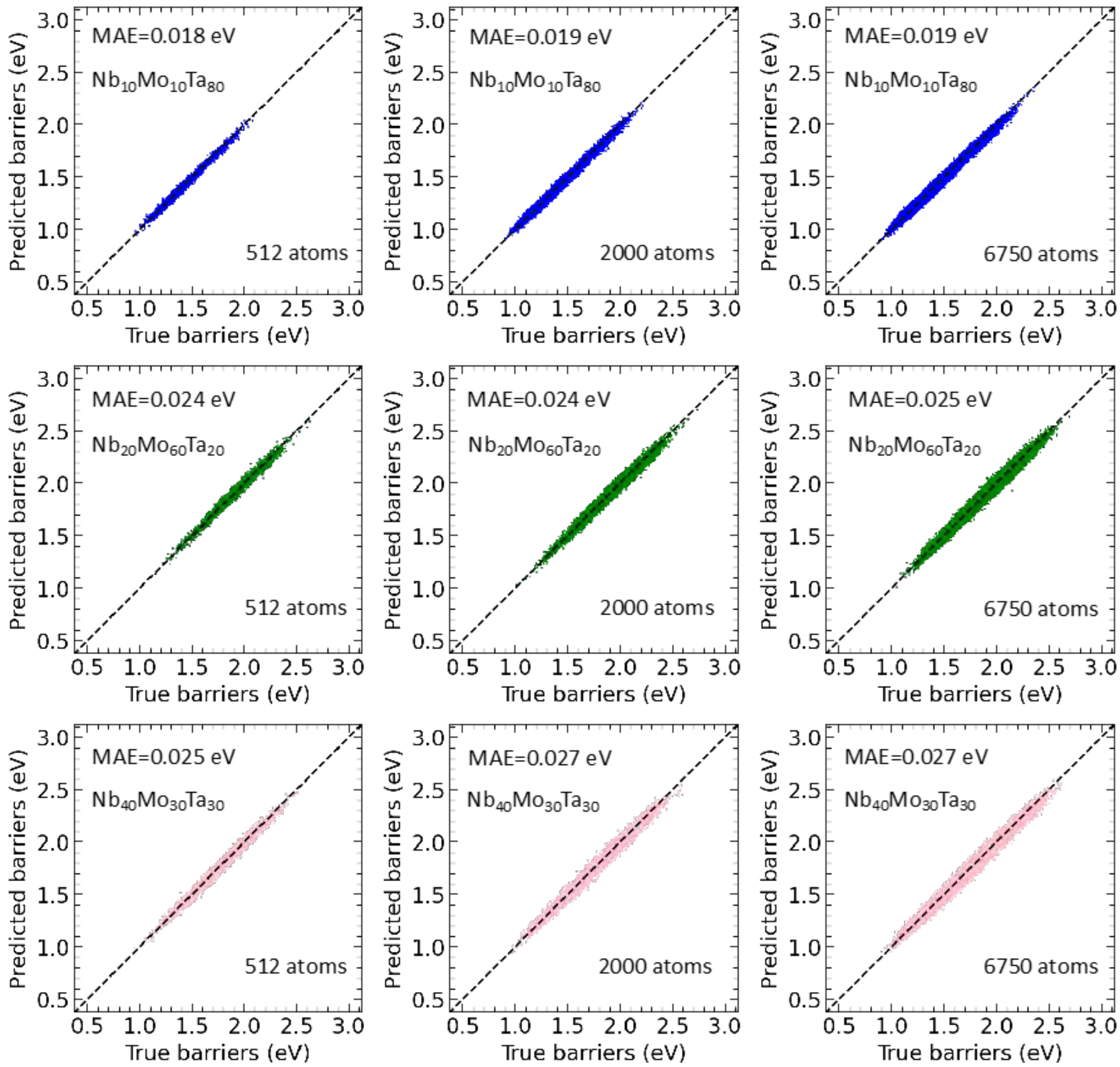

**Figure S4. Performance of neural network in predicting diffusion barrier spectrum in unseen compositions and varying system sizes (scalability).** Three compositions, including $Nb_{10}Mo_{10}Ta_{80}$, $Nb_{20}Mo_{60}Ta_{20}$, $Nb_{40}Mo_{30}Ta_{30}$, and three systems containing 512, 2000, and 6750 atoms are shown.

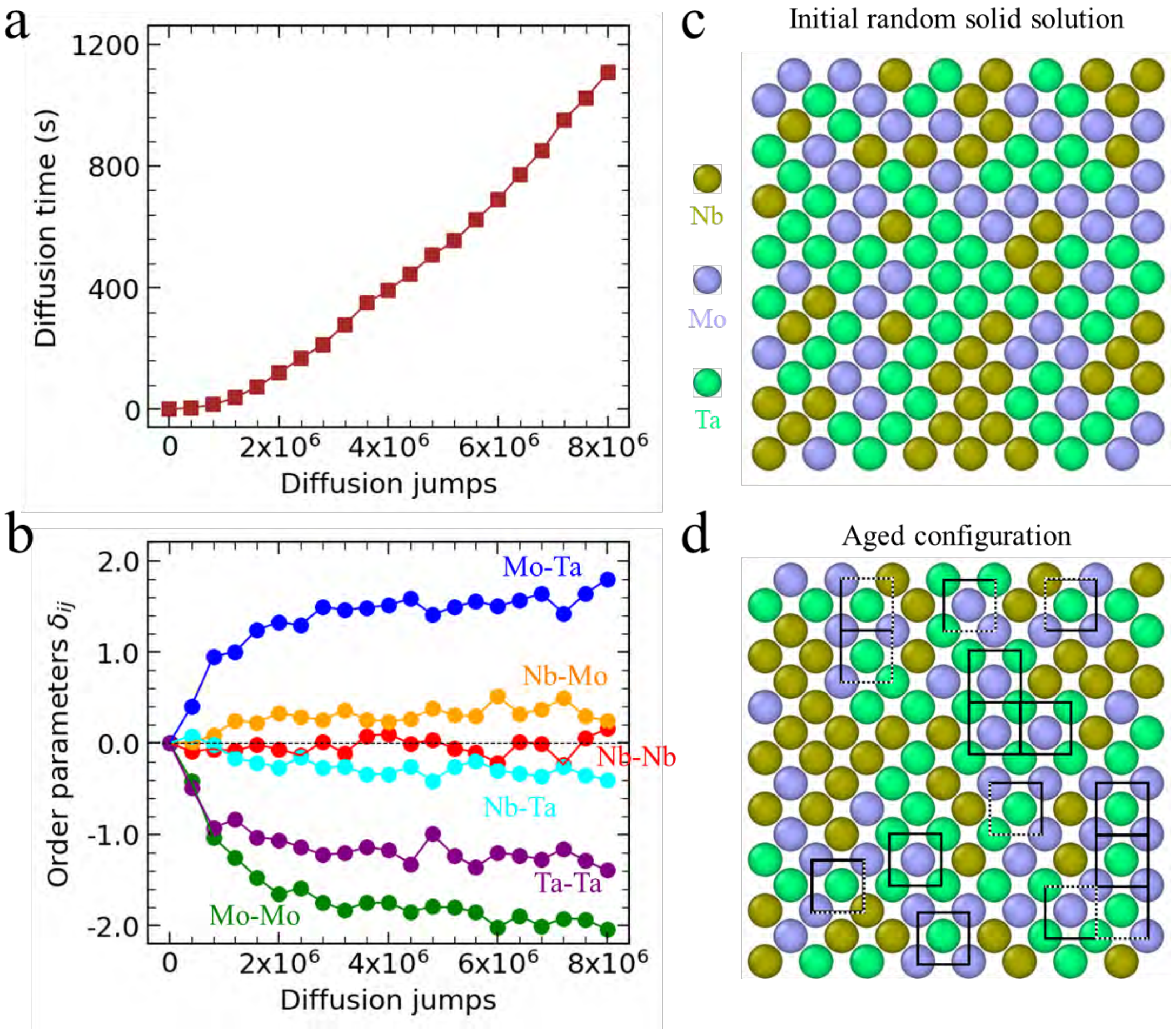

**Figure S5. Diffusion and chemical ordering in NbMoTa alloy from NNK simulation at 1,000 K.** (a) The accumulated diffusion time as a function of jumps. (b) Variation of chemical order parameters with jump. (c) Initial atomic configuration with random solid solution, and (d) aged structure demonstrating B2 ordered cluster.

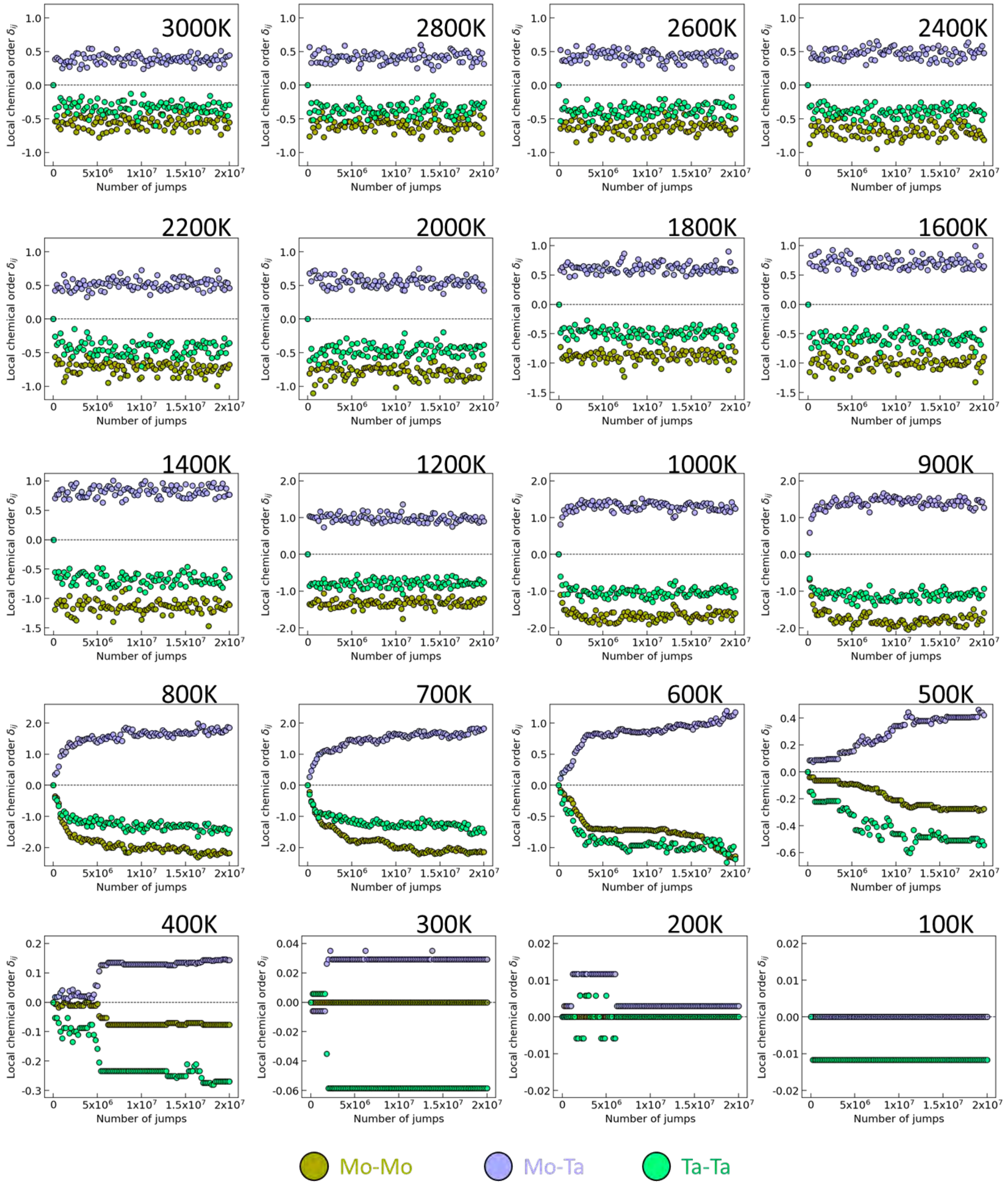

**Figure S6. Variation of the chemical order parameter as a function of diffusion jump obtained from NNK simulation.** The simulations are conducted at twenty different temperatures, ranging from 3,000 K to 100 K, as indicated in the labels.

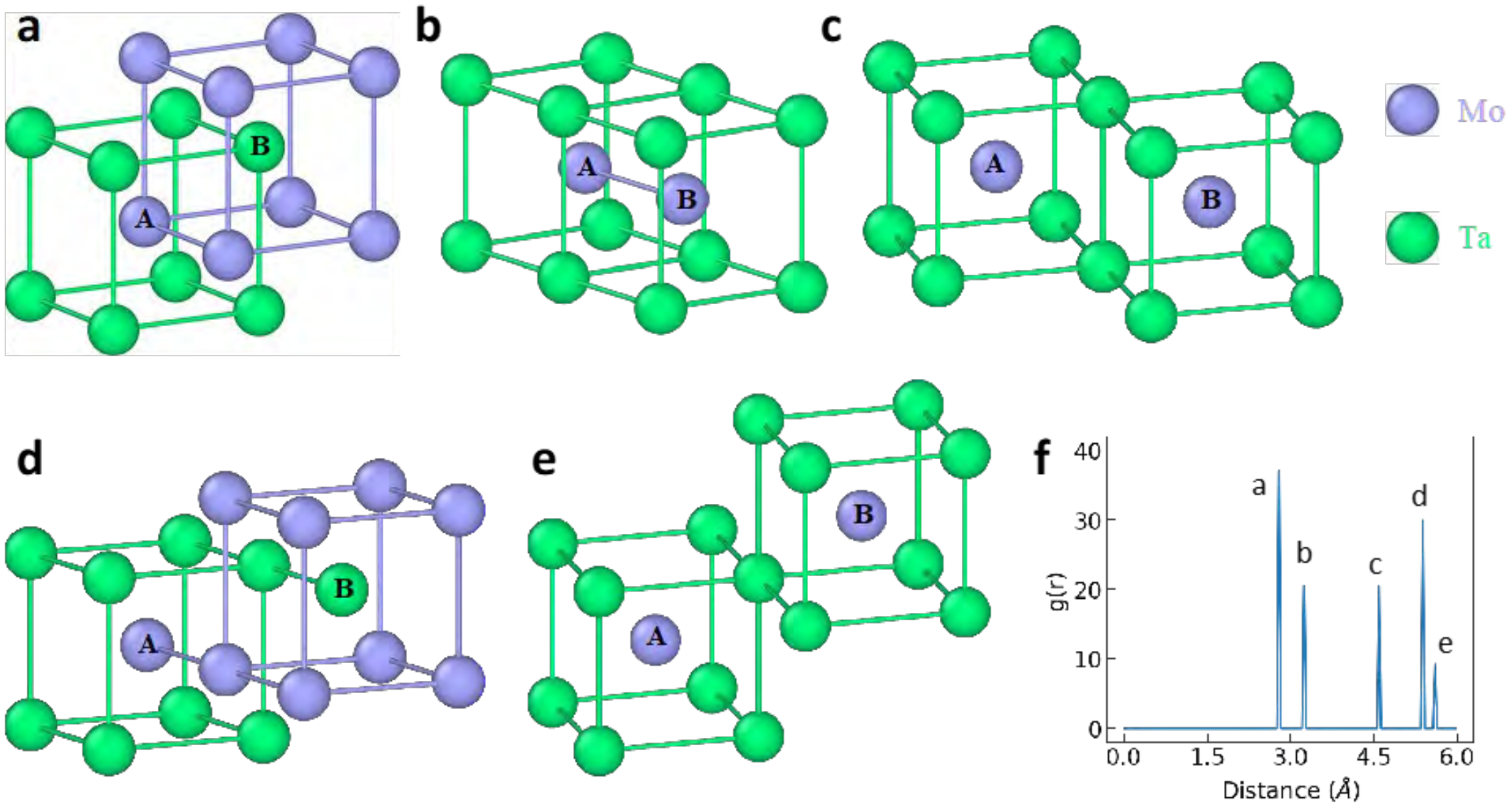

**Figure S7. B2 cluster identification.** (a-e) A cluster consists of two B2 cells that share volume, face, edge, and vertices. (f) The corresponding separation distance between the two B2 cells.

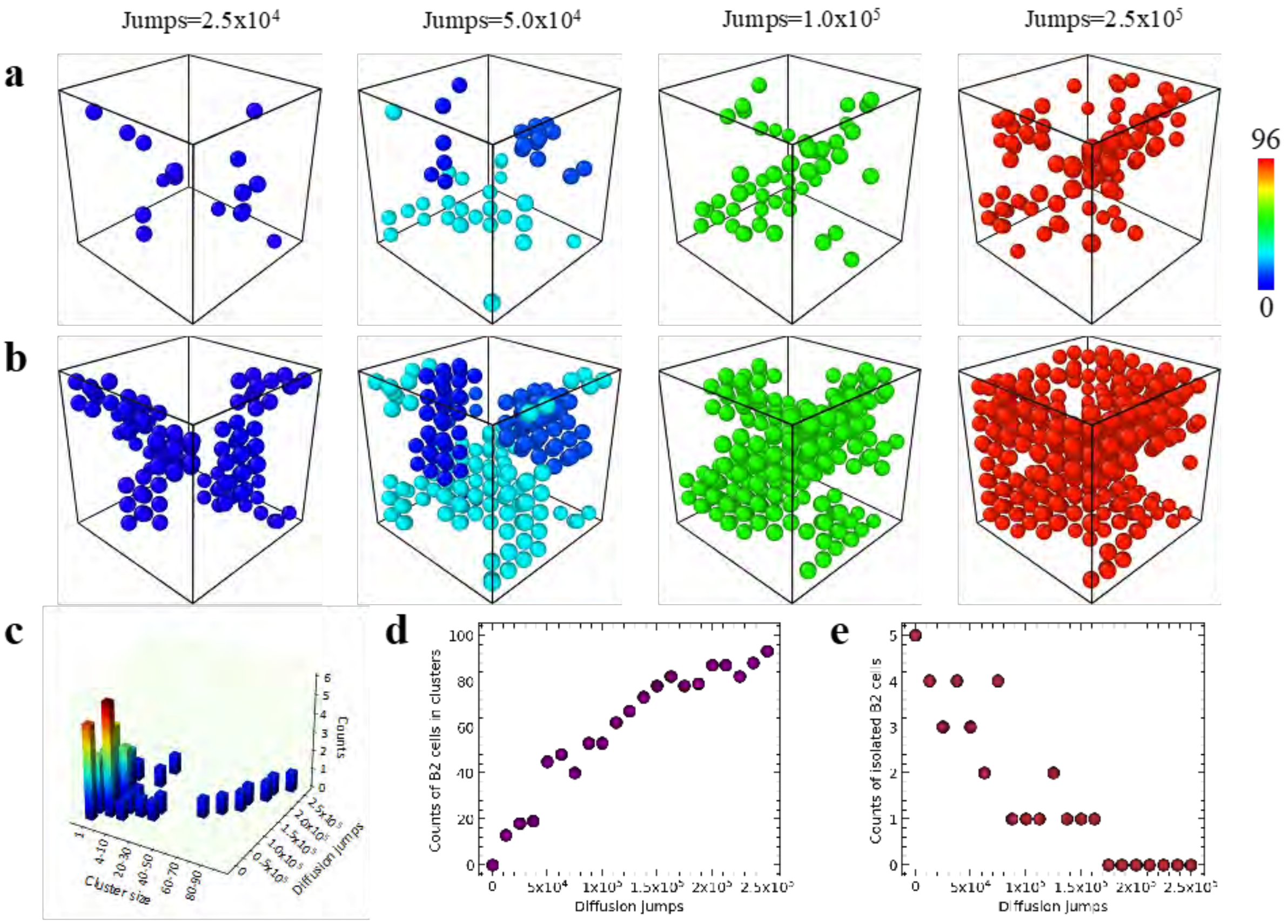

**Figure S8. Formation and coalescence of B2 clusters in a small equimolar NbMoTa model. Panel**. (a) shows the variation in B2-centred atoms as the number of jumps increases. (b) the same configuration is displayed, but with the entire B2 cells visible. After $10^5$ jumps, the two clusters combine into one, represented in green. (c) displays the B2 cluster size distribution obtained after varying numbers of jumps. (d) depicts the number of B2 cells as a function of atomic jumps, with panel (e) indicating a decrease in the number of isolated B2 cells with increasing atomic jumps.

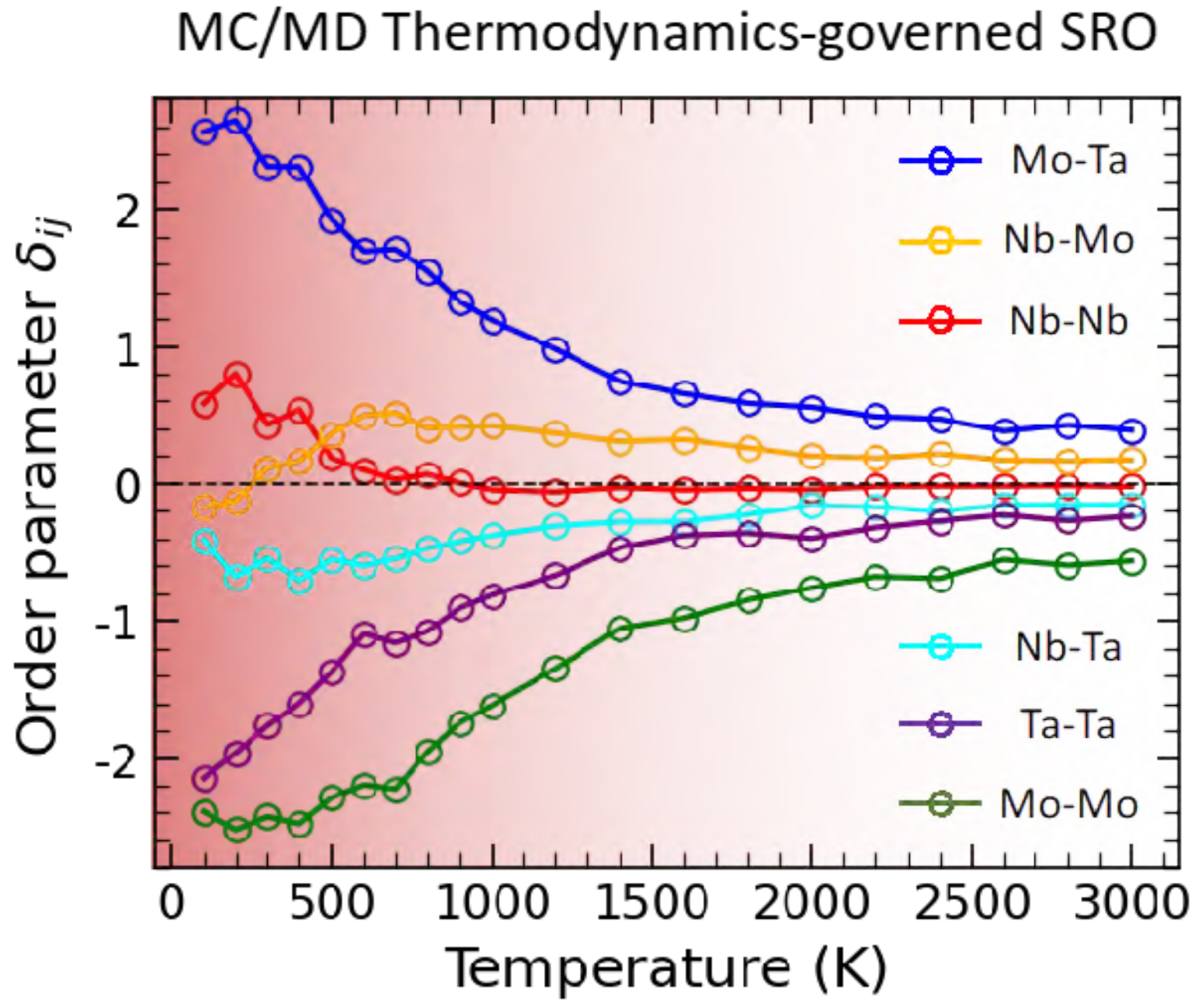

**Figure S9. Variation of chemical order obtained at different annealing temperatures using the random swap** (see Methods).

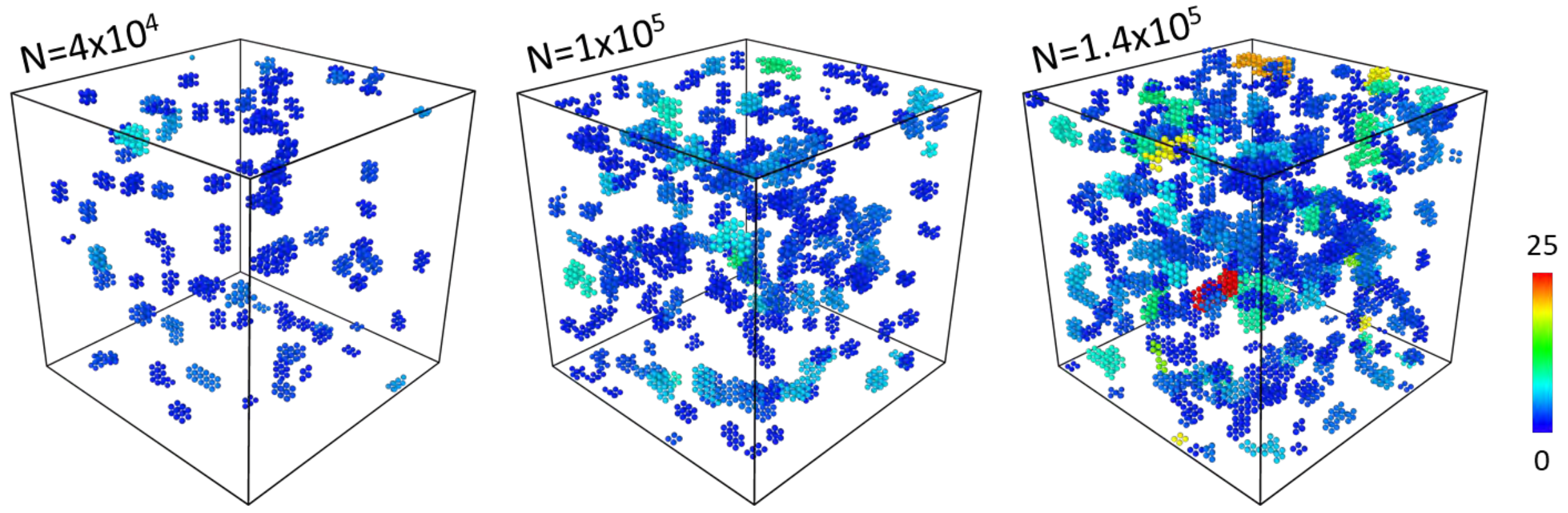

**Figure S10. B2 structure morphology generated from a random swap MCMD simulation, exhibiting a more uniform distribution.**

## 2. On-lattice structure and chemistry representation

We use on-lattice representation to convert local atomic environments into digital matrices in which each value represents one atom or vacancy. To achieve this, we follow two rules: divide the material model into a grid of pixels; place each atom at the center of one pixel. With the periodicity of crystalline structures, the rules provide us guidance in digitalizing the material model reasonably.

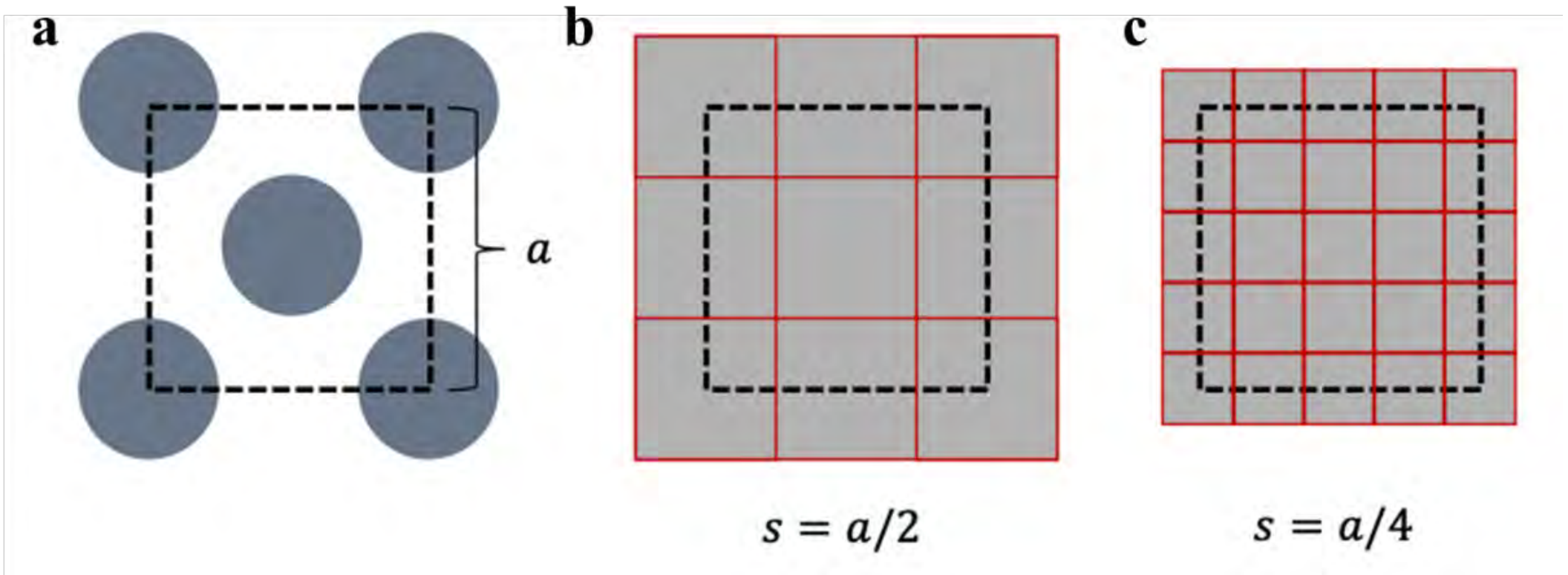

**Figure S11. On-lattice representation and pixel size determination.** (a) A unit cell with lattice constant $a$. (b-c) depicts the grid separating the atomic model into uniform cells (pixels), with the pixel size $s = a/2$ (b) and $s = a/4$ (c).

Figure S11 schematically illustrates the on-lattice representation, which converts a 2D atomic structure into a matrix. The conversion is achieved using a pixel grid, which divides the structure into uniform cells or pixels. For bcc structure, the largest grid we can use, which can fully distinct all lattices and yield the smallest voxel grid dimensionality, is $s = a/2$, where $a$ is the lattice constant of the crystal, as shown in Figure S11b. In general, the structure domain can be equally divided into pixels with size $s = a/2n$, where $n = 1, 2\ 3 \ldots$. For instance, Figure S11c shows the representation using pixel size $s = a/4$. Once converting the model into pixels, we can encode each pixel based on the local atom type as illustrated in the main text. The selection of pixel size depends on the material structure alone without involving any hyperparameters which typically exist in other structure descriptors. This avoids the need to adjust and select any hyperparameters. Furthermore, it enables us to use the largest pixel that fully captures the local structure and chemical information, reducing the burden of storage and accelerating the training of machine learning models. In Figure S1 (on page 2), we illustrate the process of converting local atomic environments into digit matrices for a 3D crystal.

**Rotational non-invariance of neural map (digital matrix).** For a given atomic configuration that includes a vacancy, there are eight migration paths associated with the vacancy in bcc crystal. The key challenge lies in how to predict these distinct migration barriers from one neural map (atomic configuration). To address this, we introduce a 'reference direction', which aims to mark the diffusion path of interest. By performing rotation and mirroring operations on the atomic configuration, we can align the diffusion direction of interest with this reference. Hence, unique digital matrices and digital vectors can be generated for each individual diffusion paths, preserving structural symmetry. Figure S12 below exemplifies this process, showing how diffusion paths 2 and 3 are aligned with the reference direction (indicted by red arrow). The Figure 2 of manuscript and Figure S3 details these operations and the resultant matrices for all diffusion directions.

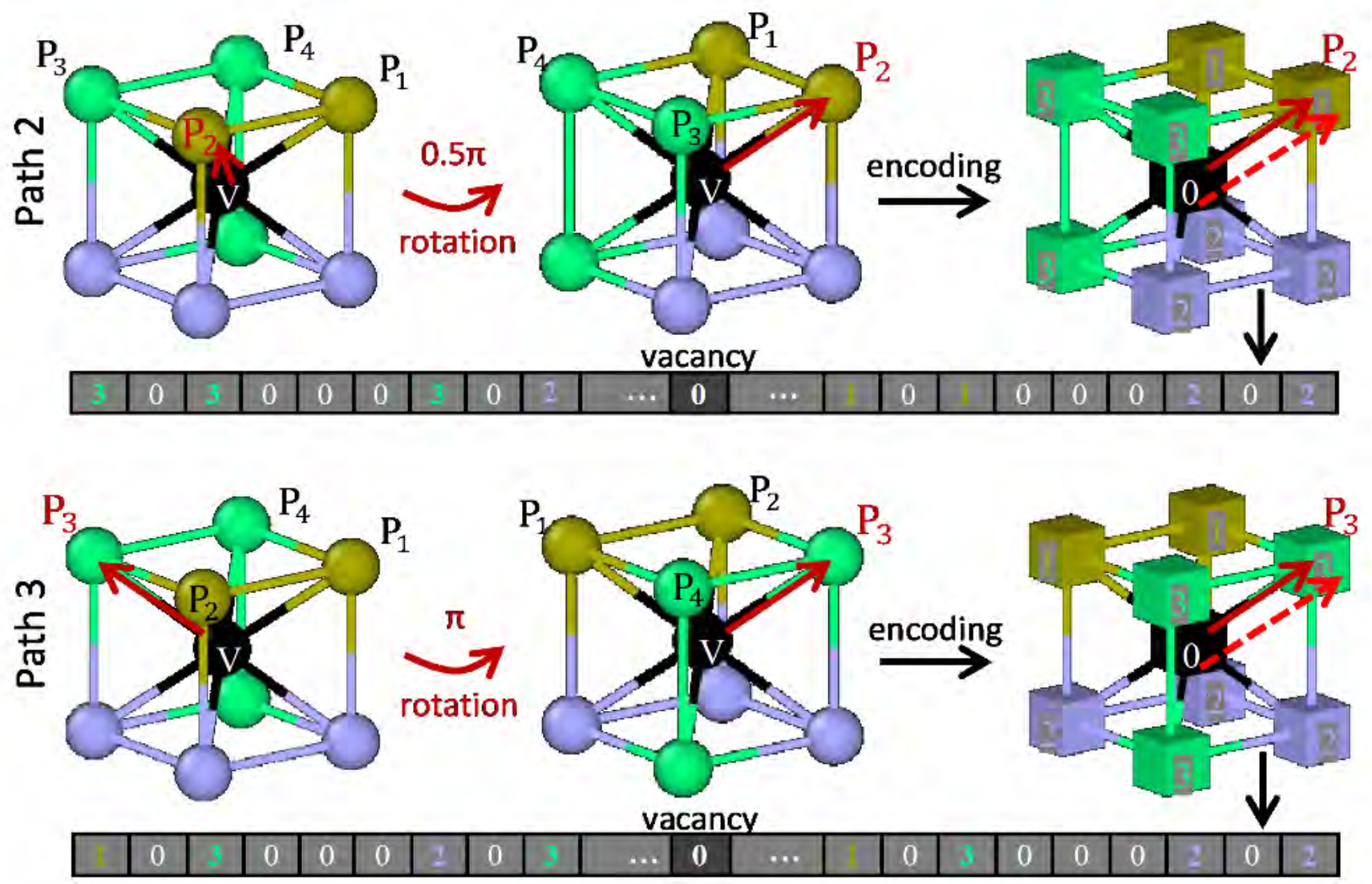

**Figure S12. Aligning diffusion paths 2 and 3 with the reference direction through rotation**. It produces two digital matrix and vectors corresponding to the two paths.

This rotational non-invariant feature of digital matrix can also be understood from the handwritten digit recognition. For instance, when the MNIST database's handwritten '6' is rotated by 180 degrees, it resembles a '9', as shown in Figure S13. Despite the pixel values in the matrices being unchanged, the orientation relative to the reference direction (denoted by the arrow) allows for the correct interpretation.

The neural network model discerns the overall sequence and pattern in the digital matrix, not individual zeros. In a perfect crystal structure (bcc here), the digital matrix displays a consistent sequence of non-zero and zero digits. However, the introduction of a vacancy alters this structure by adding an additional zero at the corresponding location. This alteration in the digital sequence is what the neural network is trained to detect and learn from, enabling it to predict associated properties.

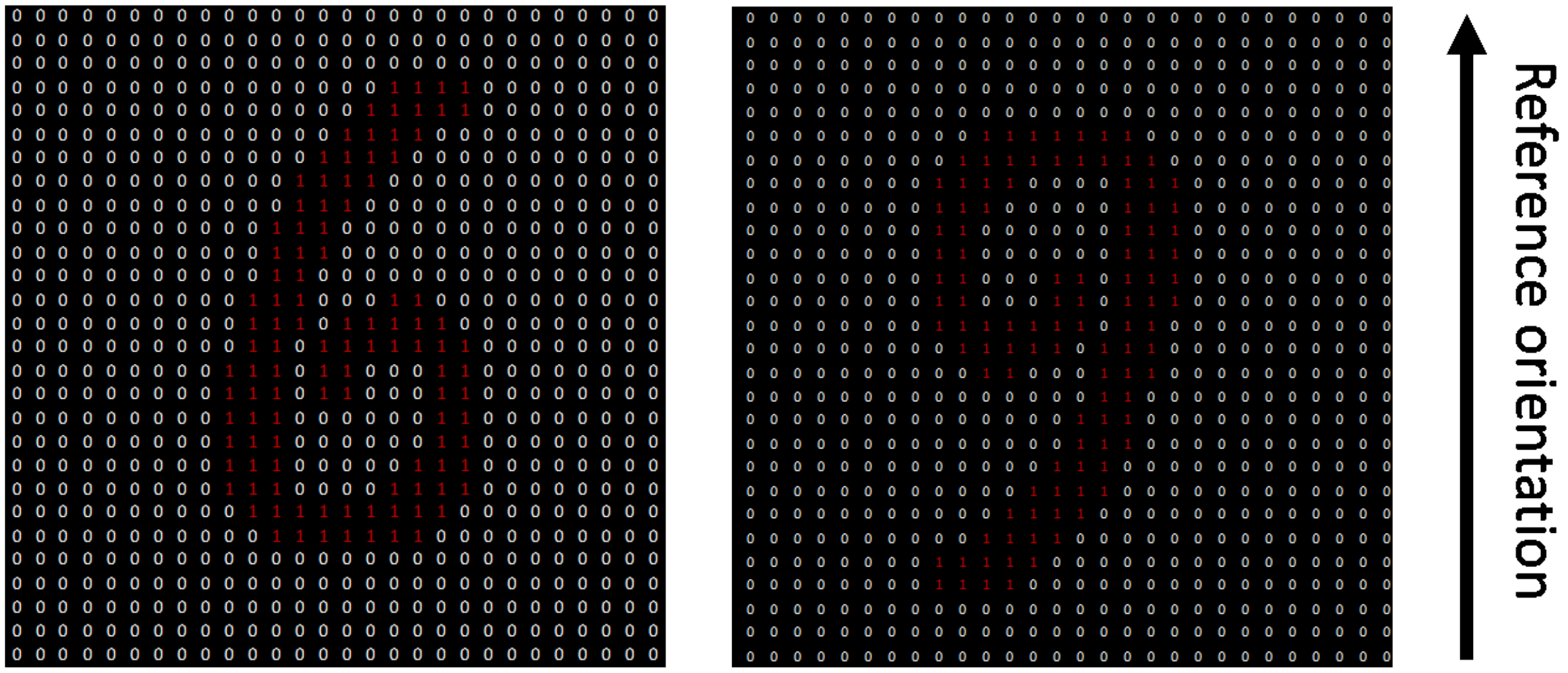

**Figure S13. Rotational non-invariance for handwritten digit recognition.**

## 3. Determining the cutoff distance

Vacancy diffusion and associated activation barrier depend on local atomic environment. The impact of surrounding atoms on vacancy diffusion should decay as the distance increases. Beyond a certain critical distance, the impact becomes negligible. To determine the critical distance, we examine the dependence of model prediction performance on the cutoff distance. Figure S14a presents a radial distribution function g(r) from a NbMoTa alloy, which indicates that atoms within 7.5 Å are separated into eight shells. When using a larger cutoff distance, we consider atoms in higher order shells, thus more atoms. Figure S14b shows the dependence of number of atoms on the cutoff distance. The number of atoms increases from 8 to 112 when the cutoff distance increases from 3.0 to 7.5 Å (meaning we consider atoms in more shells), leading to a more informative local environment representation.

For each cutoff distance, we create a dataset from four alloys, including $Nb_{33}Mo_{33}Ta_{33}$, $Nb_{50}Mo_{25}Ta_{25}$, $Nb_{25}Mo_{50}Ta_{25}$ and $Nb_{25}Mo_{25}Ta_{50}$. For each composition, we simulate atomic configurations comprising 2,000 atoms (i.e., lattice sites). Considering that each vacancy can migrate in one of eight possible directions, this results in 16,000 unique migration barriers per composition (2,000 vacancies × 8 directions). Consequently, by studying four distinct compositions, we determine a total of 64,000 barriers (16,000 barriers per composition × 4 compositions). Table S2 summarizes all the dataset. The dataset is split into two parts, with 80% used for training and 20% for validation. For each cutoff distance, we train a neural network with 4 hidden layers and 128 hidden layer units. Figure S14c shows the mean absolute errors (MAEs) of prediction on both training and validation datasets at different cutoff distances. The validation error decreases from 0.117 eV to 0.036 eV as the cutoff distance increases from 3.0 to 7.5 Å. It almost converges at later stage from 7.0 to 7.5 Å, indicating that 7.5 Å is an effective cutoff distance for representing the local atomic environment. However, we note that the neural network model and dataset have not reach a good balance for most cutoff distances, as evidenced by the gap between training and validation error. Further tuning of the network architecture can solve this problem. Nonetheless, our goal here is solely to demonstrate how the cutoff distance influences the diffusion barrier prediction using identical neural network model for all cases. We expect the conclusion will not change if we further adjust the neural network models at different cases.

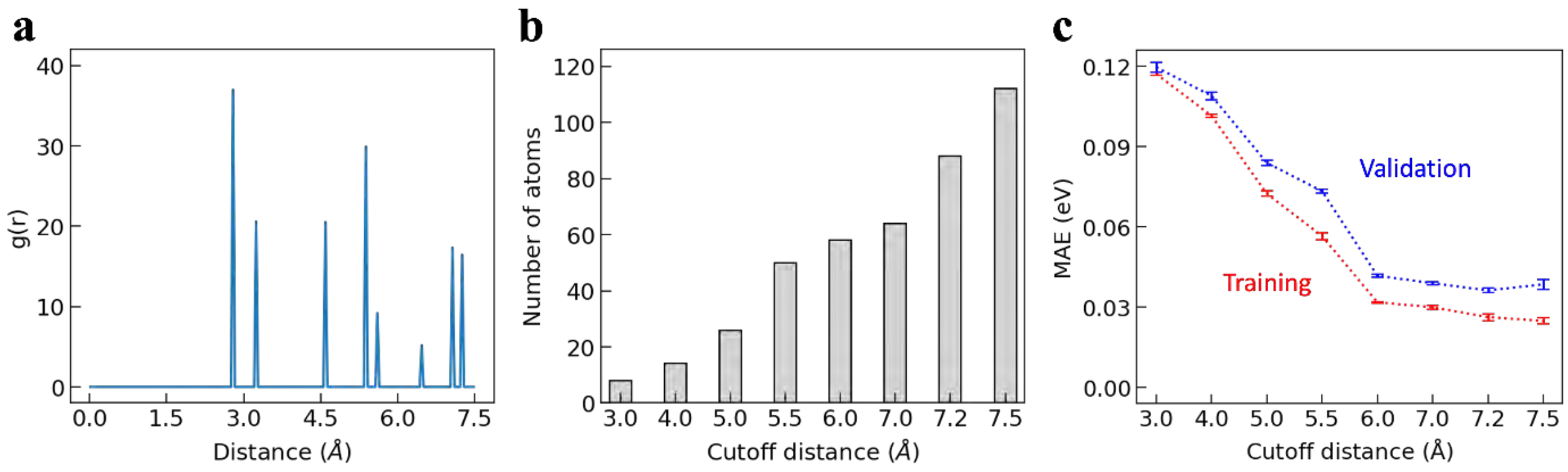

**Figure S14. Effect of cutoff distance on neural network prediction.** (a) The radial distribution function g(r) of bcc NbMoTa. (b) The number of neighboring atoms surrounding a vacancy as a function of cutoff distance. (c) The machine learning prediction error as a function of cutoff distance has converged at 7.5 Å. The red and blue curve represents the training and validation mean absolute error (MAE), respectively. The error bars represent the standard deviations of model prediction errors using five-fold cross-validation.

## 4. Architecture of neural network and convolutional neural network

The two critical parameters determining the architecture of a neural network include the number of layers, the number of neurons in each layer. To understand the influence of architecture on prediction performance, we train different neural networks using a dataset containing 46 compositions. We compute and generate 736,000 vacancy barriers from these 46 compositions (16,000 barriers from each composition), and the Table S3 summarizes the compositions and dataset. The dataset is split into two parts, 95% as training dataset and 5% as validation dataset. We train a set of neural networks with different numbers of hidden layers (from 1 - 4) and numbers of hidden layer units (16 - 256). We use 69,920 data points (10% of the whole training dataset) to train the networks and then compare the performance of different models on the validation dataset. Figure S15b shows the mean absolute errors of prediction for these models, and Figure S16 presents a direct comparison between true values and predicted value from different neural network models. The prediction error decreases with either increasing the number of layers or neurons and begin to converge for the model with 128 neurons and 2 layers. This suggests that the second order interaction from two hidden layers is sufficient to capture the vacancy-atom interactions. Additionally, the convergence on 128 neurons has physical meaning as they can explicitly capture the 112 neighboring atoms of a vacancy.

In addition to the classic neural network, we have also trained a convolutional neural network (CNN) using the same datasets. Figure S17 depicts the structure of the CNN, which comprises one input layer, four convolutional layers, and one output layer. To the input layer, we feed the 3D neuron map (images), and in each of the four convolutional layers, we apply filters of size $3 \times 3 \times 3$. The number of filters used in the convolutional layers is 32, 64, 128, and 128, which is equivalent to the number of channels of the generated images. Consequently, the data dimension reduces to $1 \times 1 \times 1 \times 128$ from the original $9 \times 9 \times 9 \times 1$. Following each convolutional operation, we apply batch normalization (before the activation function), which provides benefits such as a reduction of sensitivity to model parameter initialization, regularization. The Rectified Linear Unit (ReLU) serves as the activation function, and the data from the final convolutional layer is converted to a one-dimensional vector of length 128 before being passed to the output layer. The output layer, comprising a single neuron, predicts the diffusion barrier. The CNN model is trained for 100 epochs using the Adam optimizer with an initial learning rate of 0.001 and a batch size of 32. After each epoch, the model is evaluated on the validation dataset to monitor the evolution of the loss, which is represented by the mean square error. If the validation loss fails to decrease after 10 consecutive epochs, the learning rate is decreased by a factor of 10. This smaller learning rate reduces oscillation, avoids divergence of the optimization, and contributes to convergence to the nearby minimum point. The minimum allowed learning rate is $1 \times 10^{-5}$, as a learning rate that is too low can greatly slow down the training procedure and waste computational resources. Once the learning rate reaches the minimum value, it remains constant for the rest of the training process. The training procedure ends after 100 epochs, regardless of whether the learning rate has reached the minimum value. The parameters of the model with the best performance are saved for further use.

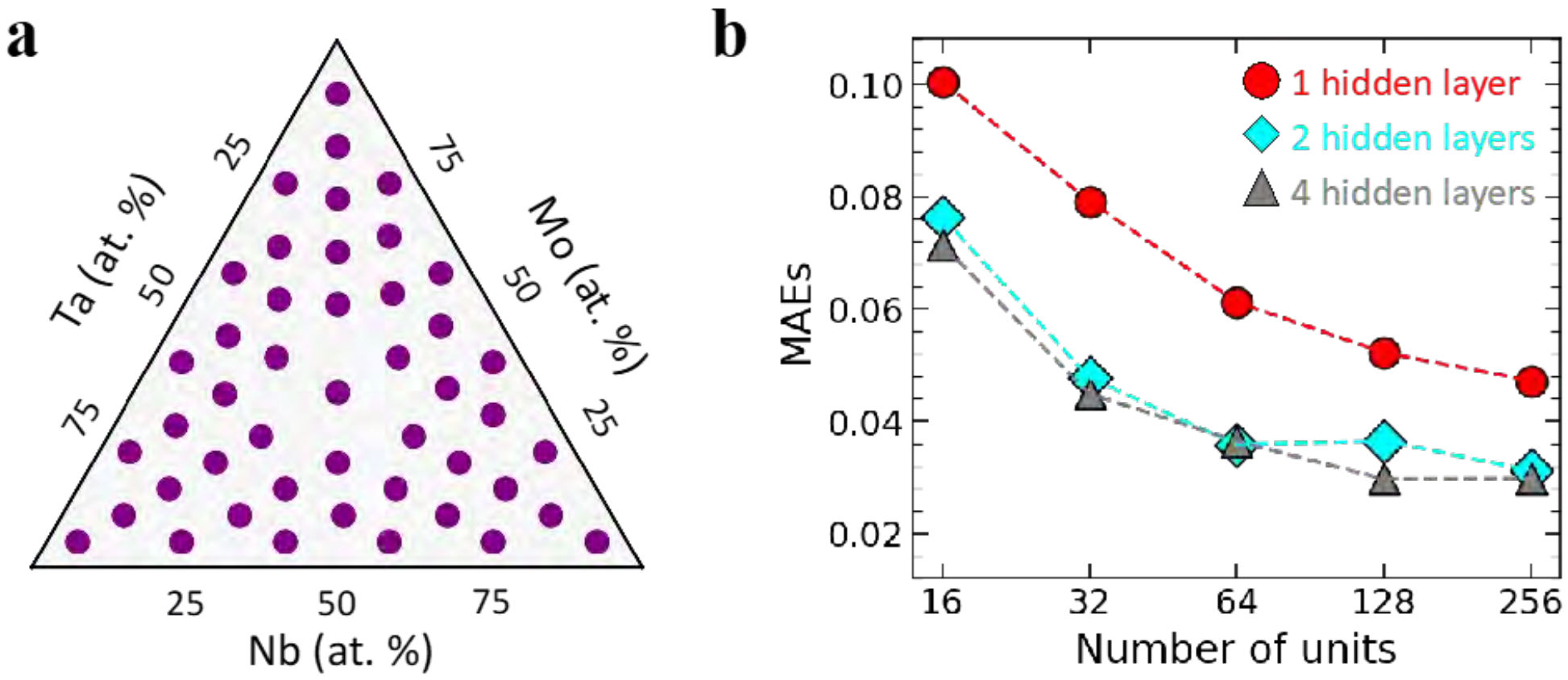

**Figure S15. Neural network prediction performance with different numbers of hidden layers and units.** (a) For the neural network model, 46 compositions selected from the NbMoTa compositional space for training. (b) Different neural network models are evaluated based on their prediction errors on the validation dataset.

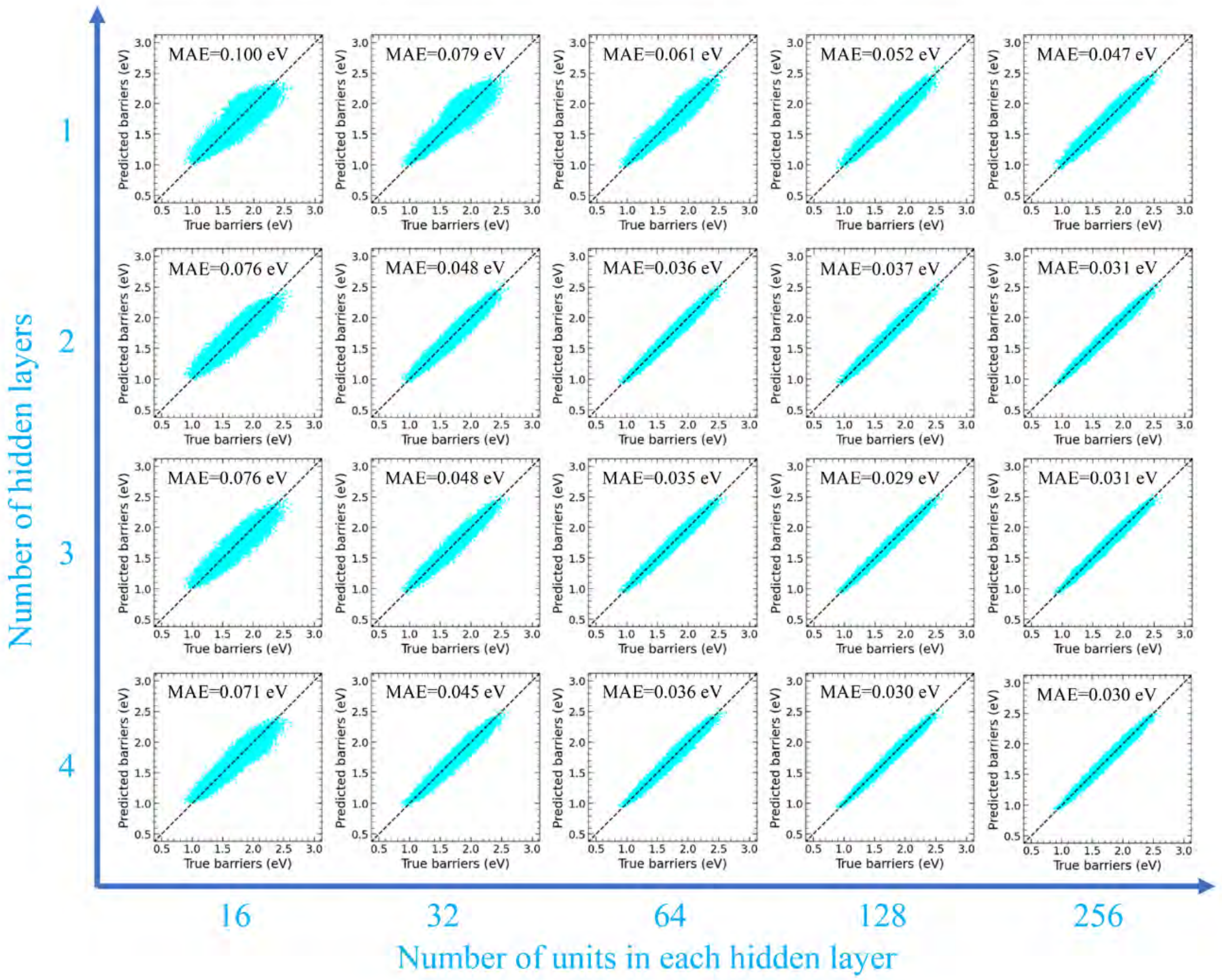

**Figure S16. Prediction performance of neural network models with varying hidden layers and the number of units.** The prediction accuracy increases with increasing the number of hidden layers and units.

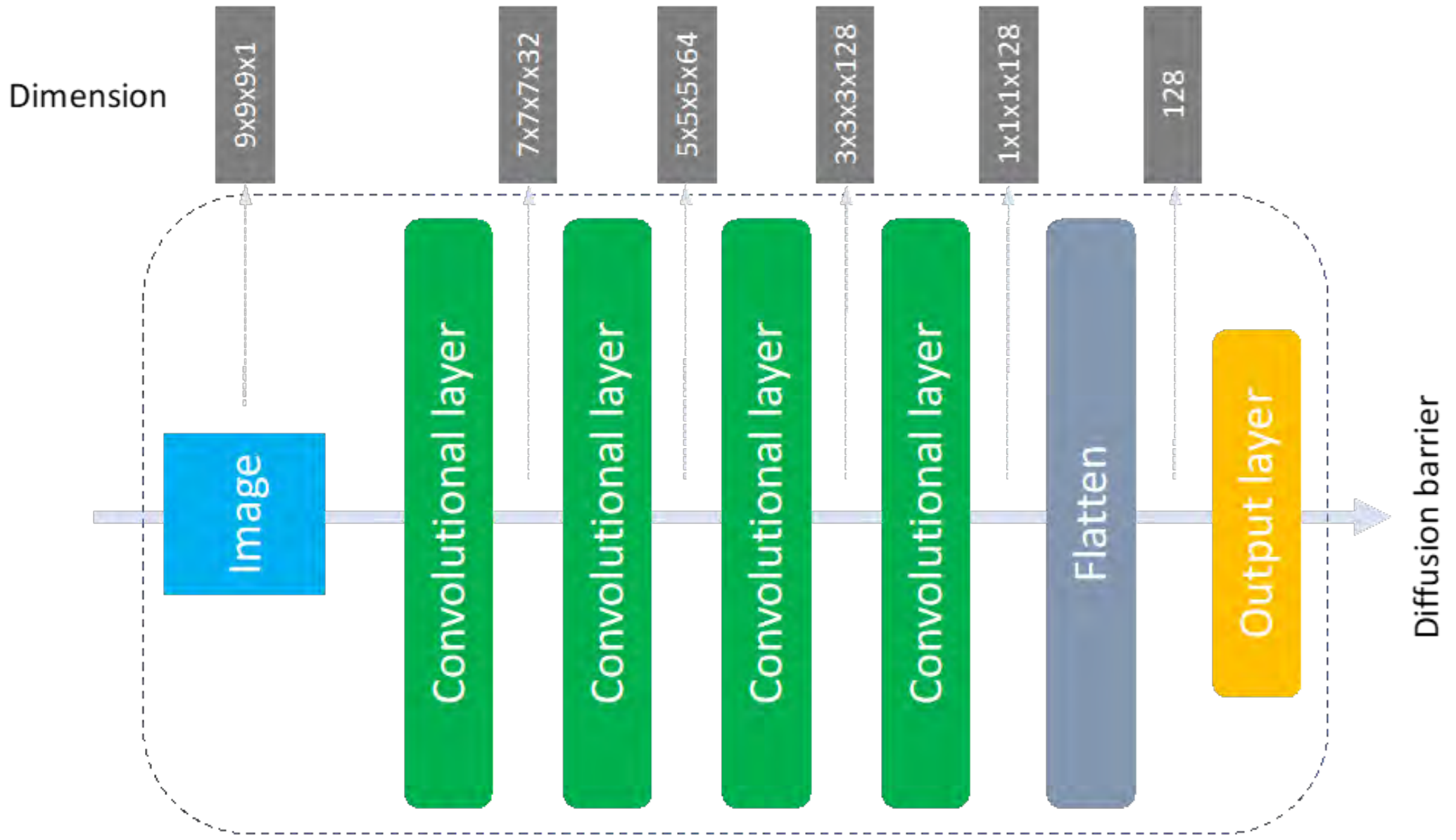

**Figure S17. Architecture of the convolutional neural network, consisting of one input layer, four convolutional layers, and one output layer.**

## 5. Number of compositions for predicting the entire ternary composition space

We select different numbers of compositions to train both neutral network and CNN models, in order to understand how many compositions are required to predict the entire compositional space. Figure S18a depicts the 46 compositions uniformly distributed in the compositional space. Figure S18b-d, illustrate 1, 4 and 10 compositions (red colored points) located at the center region of the compositional space, respectively. Each composition comprises 16000 barrier data points. For 1-composition dataset, 80% and 20% data are used for training and validation respectively. For 4-composition dataset, 90% and 10% data are used for training and validation respectively. For 10-composition and 46-composition datasets, 95% and 5% data are used for training and validation, respectively.

Figure S19 presents the prediction performance (i.e., MAEs) in an unseen equimolar NbMoTa alloy as function of number of compositions used for training models. The prediction error decreases rapidly when the number of training compositions increases from 1 to 4. When the number of compositions increases from 4 to 46, the prediction error is further lowered with a small amount. The trend indicates that the addition of data from dilute solutions (i.e., the corners of compositional space) can improve model prediction performance, but not as significant as concentrated solutions. The CNN model performance from 4-composition dataset (MAE = 0.026 eV) is remarkable (the average barrier is around 1.5 eV), which implies that the CNN model has deciphered the chemical complexity and successfully linked it with diffusion barriers. As to the neural network models, it is worthwhile to note that the prediction errors barely change when we increase the number of hidden layers, suggesting the 4 layers of neural network are sufficient for the barrier prediction. Compared to neural network, the CNN shows enhanced performance with lower MAEs, implying the added convolutional layers capture the large-scale atomic patterns contributing to vacancy migration.

It is worth nothing that the testing performance of the trained neural networks is evaluated using newly generated data from other (unseen) compositions, which are not used for training and validation. Figure S20 shows the testing results for these new compositions, $Nb_{10}Mo_{10}Ta_{80}$, $Nb_{20}Mo_{60}Ta_{20}$, $Nb_{40}Mo_{30}Ta_{30}$, and the average MAE is smaller than 0.018 eV, implying the generalizability. Notably the neural networks trained using small configurations precisely predict diffusion barriers in large atomic configuration. For instance, the neural network preserves the consistent high accuracy for different sized systems containing 512, 2000, and 6750 atoms, indicating scalability.

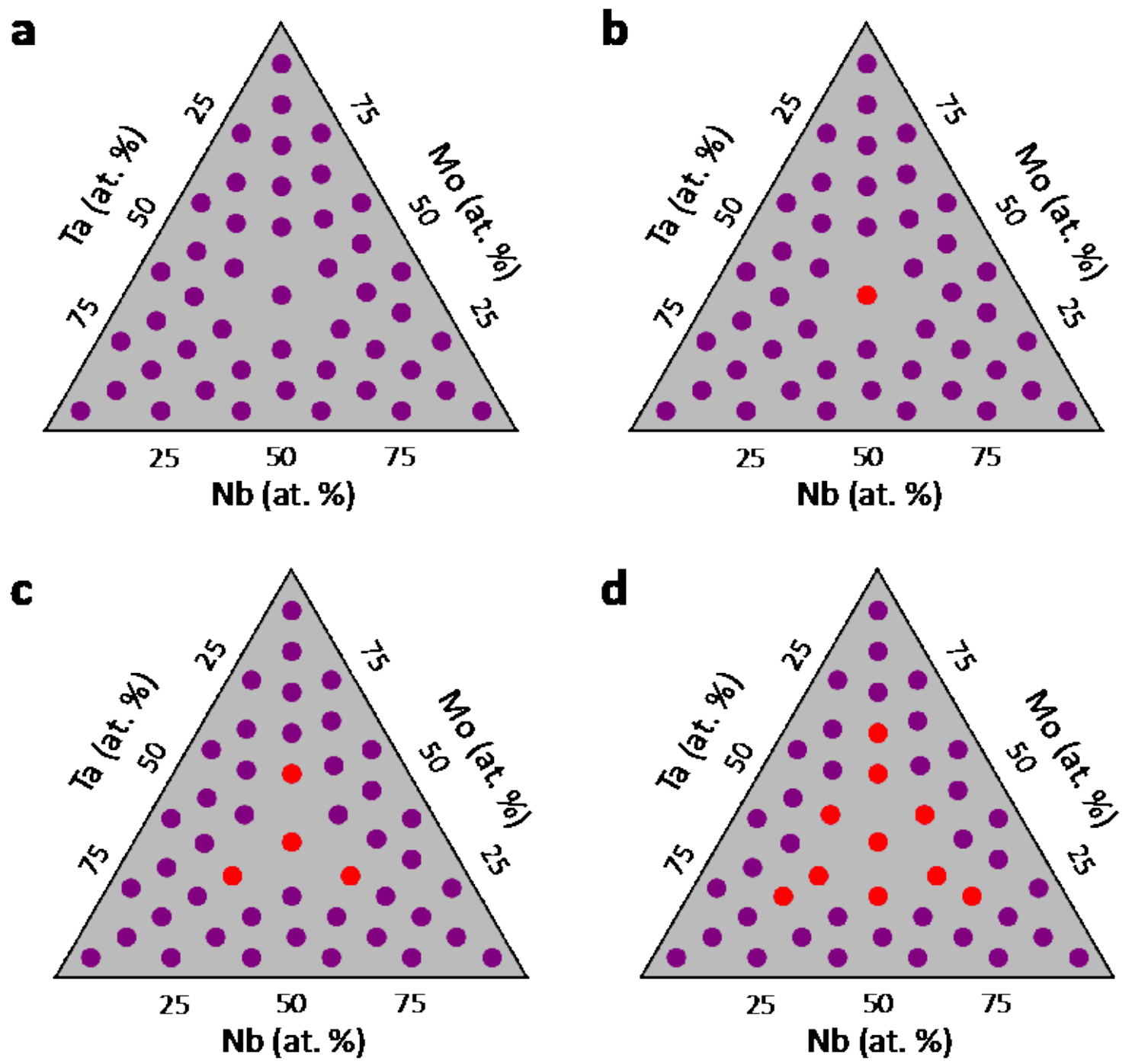

**Figure S18. Compositions used for building different training datasets.** (a) depicts forty-six compositions occupying the NbMoTa compositional space uniformly. (b-d) depict one composition, four compositions, ten compositions (red-colored points) located at the center of the compositional space representing concentrated alloys.

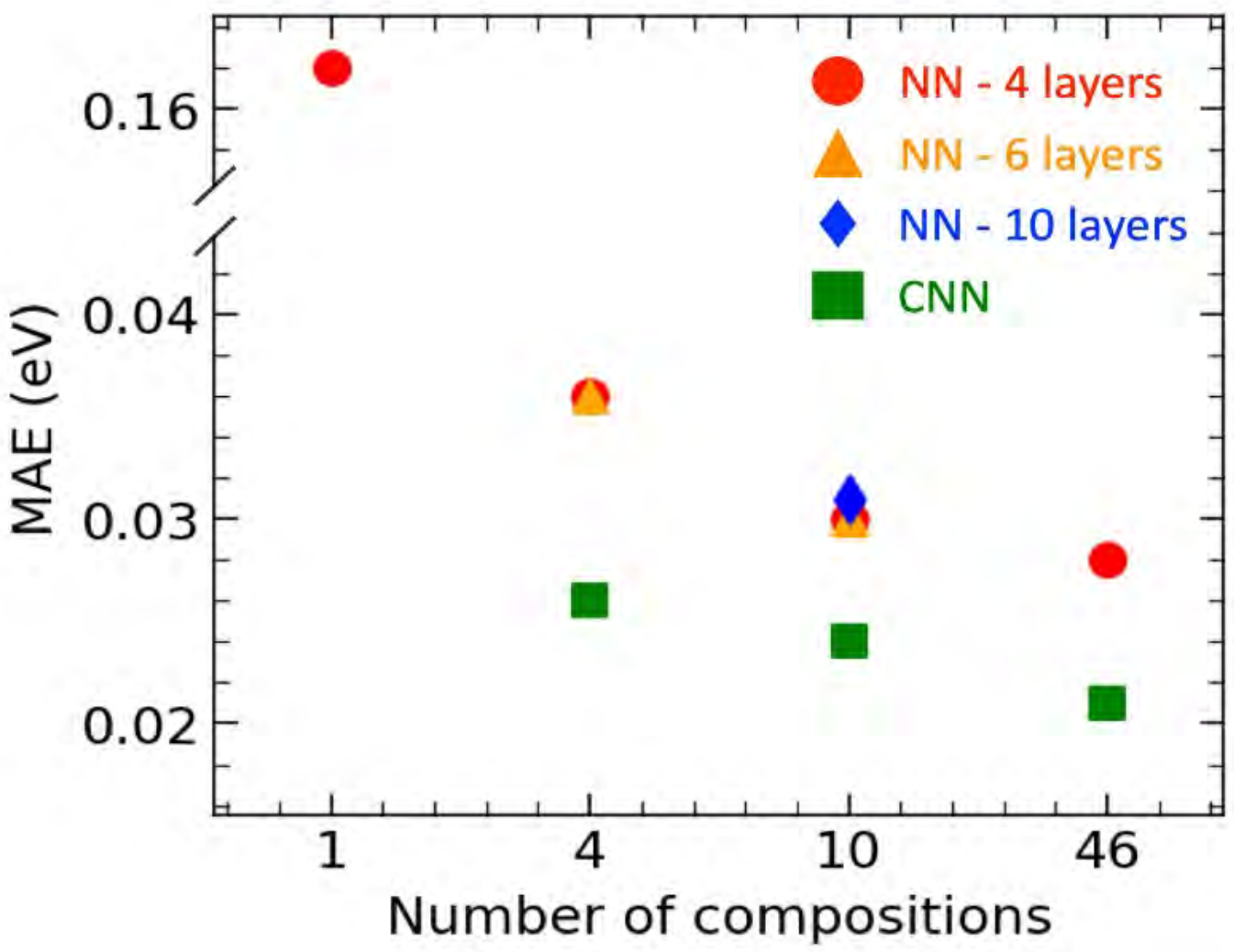

**Figure S19. Prediction error of neural network (NN) and CNN as a function of the number of compositions used during training.** The evaluation is done on previously unseen compositions in Nb-Mo-Ta, and the results indicate that including more than four compositions leads to a rapid convergence of the network's performance.

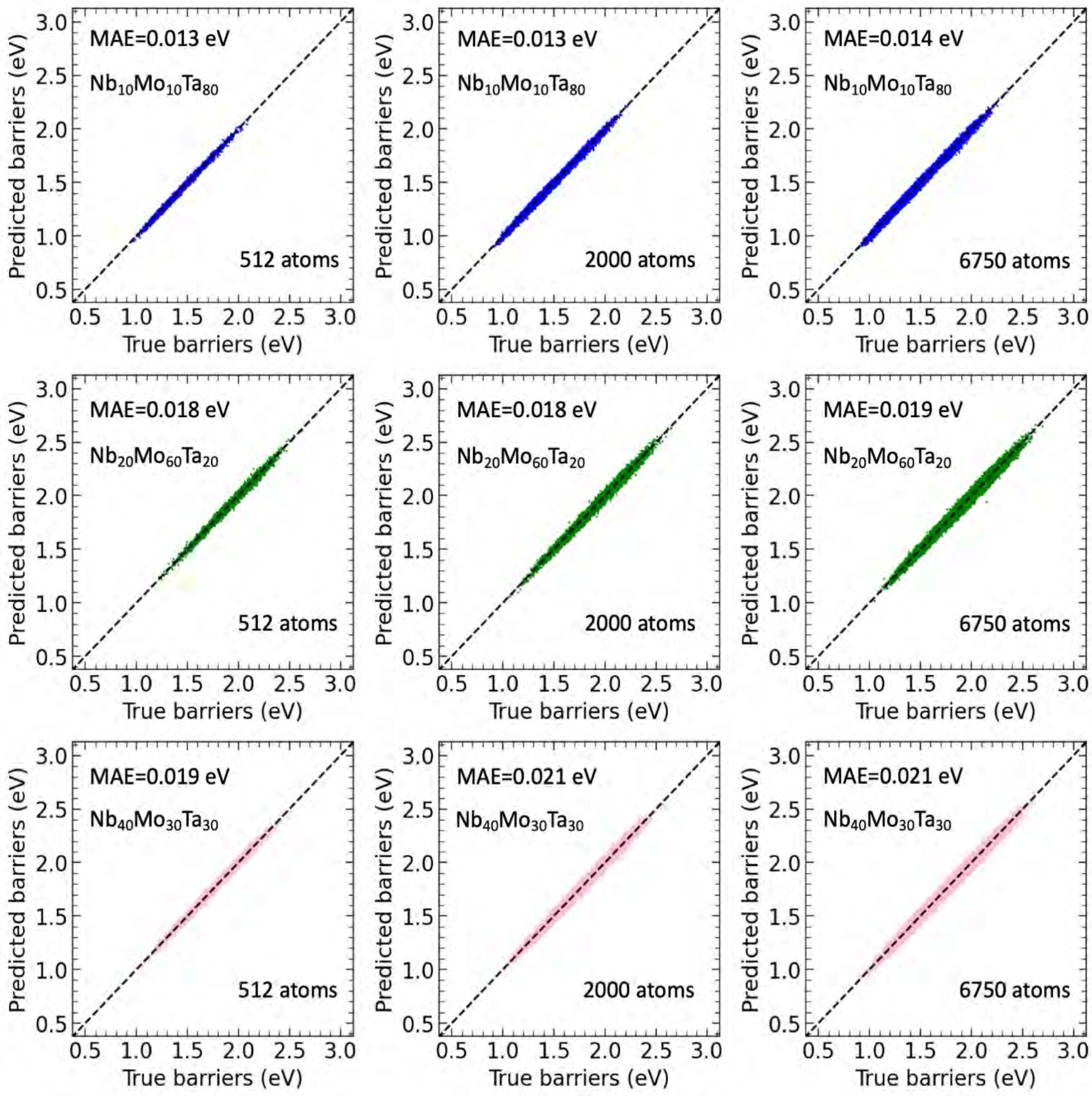

**Figure S20. Performance of CNN in predicting diffusion barrier spectrum in unseen compositions and varying system sizes (scalability).** Three compositions, including $Nb_{10}Mo_{10}Ta_{80}$, $Nb_{20}Mo_{60}Ta_{20}$, $Nb_{40}Mo_{30}Ta_{30}$, and three systems containing 512, 2000, and 6750 atoms are shown. The architecture of Convolutional neural network is illustrated in Figure S14.

## 6. Comparison with cluster expansion method

The cluster expansion (CE)[1,2] method is often used to predict thermodynamic properties of multicomponent systems, such as vacancy formation energy[3]. The total energy of a configuration is computed by summing up a series of clusters (for instance, single, pair, triplet, and large group of atoms). For kinetics problems such as vacancy diffusion, the governing property is the underlying diffusion activation energy $\Delta E$, i.e., the energy difference (barrier) between transition state and the initial energy minimum. Figure S1 schematically illustrate vacancy diffusion and its corresponding potential energy landscape. The process starts from an initial state $E_i$, through a transition state $E_s$, and leads to the neighboring local minimum, i.e., final state $E_f$. The CE has been adopted to predict the energies of local minimum states. However, addressing the transition state presents specific challenges in CE method.

In the classic application of CE for predicting configurational energy, the initial critical steps involve designing unique clusters based on lattice symmetry and choosing an optimal cluster set, which can be a time-consuming process. For example, the property of vacancy can be influenced by the atoms in its 8$^{th}$ neighboring shell. It has been noted that it can take a number of weeks to months to select optimal clusters when the 8$^{th}$ nearest neighboring atoms is considered [4]. There have been attempts to use CE method for predicting saddle point energy in binary alloys[4], and to our best of knowledge, only one study[5] focusing on predicting vacancy barriers in compositionally complex alloys (ternary alloys). There are notable challenges associated with employing CE for diffusion barrier prediction.

(i) Saddle point representation. To model the transition state (saddle point) for energy prediction using CE, the strategy is to introduce artificial atoms. In a binary system, the jumping species could be either atom type 1 or 2. To differentiate, two additional species, type 3 and type 4, are introduced. This increases chemical complexity and the intricacy of cluster design. In a N-component system, N extra species need to be defined, leading to a total of 2N+1 species (including vacancy).

(ii) CE prediction performance in ternary alloy. To predict vacancy diffusion barrier in ternary system (Al-Mg-Zn), the CE is combined with a quartic function of the reaction coordinate[5]. Comparing with the predicted diffusion barriers with the ground truth obtained from NEB calculation, the mean error of CE is 0.0451 eV, approximately 10% of the average actual barrier. This is contrasted with our model, as shown in Figure 2 of the manuscript, which achieves a significantly lower mean error of 1.2%, an order of magnitude lower than that of CE.

We discuss some key features associated with our neural network kinetics model, which render its highly accurate barrier prediction. It is noted that the model performance in highly accurate barrier prediction is not just for a single alloy composition but across a wide range of varying alloy compositions (the entire compositional space of the ternary alloys).

a. Neuron map representation of atomic structure and chemistry: The neuron map (on-lattice) representation precisely captures atomic structure and composition. Its dimension $O(N)$ scales linearly with the number of atoms $N$, and has the lowest dimensionality possible as a crystal descriptor. Critically, determining the neuron map is simple and involves simple calculation (avoiding the painstaking parameter tuning in other methods, such as cluster design in CE).

b. Predict performance: The model exhibits high accuracy in barrier prediction. For instance, the mean absolute errors (MAE) for dilute solution $Nb_{90}Mo_5Ta_5$ and concentrated solution $Nb_{33}Mo_{33}Ta_{33}$ are 0.011 and 0.021 eV, respectively. The error is smaller than 1.2% of the true diffusion barrier.

c. Generalization and predicting in entire compositional space: More importantly for compositionally complex materials possessing a vast compositional space, the current method, trained on dozens of compositions, shows remarkable predictability for new (previously unseen) compositions, allowing accurate mapping of the entire ternary space (Figure 2c of manuscript).

d. Scalability and efficiency: Our model demonstrates scalability with system size. This size scalability is shown by accurate barrier predictions in larger NbMoTa systems. For example, the neural network preserves a consistent high accuracy for different sized systems containing 512, 2000, and 6750 atoms.

e. High efficiency in modeling diffusion: neural map originates from its simplicity to mirror vacancy jumps through the swapping of neurons (digits). With only one-time conversion of atomic configuration to neuron map, vacancy jumps and chemical evolution can be simulated by swapping two digits of neural map. In this way, millions of vacancy jumps can be modeled efficiently, with each jump iteration involving the action of just two neurons. Using one single CPU, the model evolves 10 million diffusion jumps in a large system containing 128,000 atoms within two days.

Compared with traditional CE, our introduced neuron map representation and the computational scheme are more accurate (low error prediction), earlier to and capable of predicting vacancy barriers in the entire space using small training data (Figure 2c of the manuscript), and easy and fast (Figure 5, evolving 10 million jumps in large system).

## 7. Supplementary Tables

**Table S2. Dataset for determining the cutoff distance**

| Index | Composition | System size (atoms) | Number of barriers (NEB calculation) | Total number of barriers |
|---|---|---|---|---|
| 1 | $Nb_{33}Mo_{33}Ta_{33}$ | 2,000 | 16,000 | 64,000 |
| 2 | $Nb_{50}Mo_{25}Ta_{25}$ | 2,000 | 16,000 | |
| 3 | $Nb_{25}Mo_{50}Ta_{25}$ | 2,000 | 16,000 | |
| 4 | $Nb_{25}Mo_{25}Ta_{50}$ | 2,000 | 16,000 | |

**Table S3. Dataset used for predicting vacancy diffusion barriers in the entire Nb-Mo-Ta space**

| Index | Composition | System size (atoms) | Number of barriers (NEB calculation) | Total number of barriers |
|---|---|---|---|---|
| 1 | $Nb_{5}Mo_{5}Ta_{90}$ | 2,000 | 16,000 | 736,000 |
| 2 | $Nb_{5}Mo_{22}Ta_{73}$ | 2,000 | 16,000 | |
| 3 | $Nb_{5}Mo_{39}Ta_{56}$ | 2,000 | 16,000 | |
| 4 | $Nb_{5}Mo_{56}Ta_{39}$ | 2,000 | 16,000 | |
| 5 | $Nb_{5}Mo_{73}Ta_{22}$ | 2,000 | 16,000 | |
| 6 | $Nb_{10}Mo_{10}Ta_{80}$ | 2,000 | 16,000 | |
| 7 | $Nb_{10}Mo_{27}Ta_{63}$ | 2,000 | 16,000 | |
| 8 | $Nb_{10}Mo_{44}Ta_{46}$ | 2,000 | 16,000 | |
| 9 | $Nb_{10}Mo_{61}Ta_{29}$ | 2,000 | 16,000 | |
| 10 | $Nb_{15}Mo_{15}Ta_{70}$ | 2,000 | 16,000 | |
| 11 | $Nb_{15}Mo_{33}Ta_{52}$ | 2,000 | 16,000 | |
| 12 | $Nb_{15}Mo_{51}Ta_{34}$ | 2,000 | 16,000 | |
| 13 | $Nb_{20}Mo_{20}Ta_{60}$ | 2,000 | 16,000 | |
| 14 | $Nb_{20}Mo_{40}Ta_{40}$ | 2,000 | 16,000 | |
| 15 | $Nb_{5}Mo_{90}Ta_{5}$ | 2,000 | 16,000 | |
| 16 | $Nb_{22}Mo_{73}Ta_{5}$ | 2,000 | 16,000 | |
| 17 | $Nb_{39}Mo_{56}Ta_{5}$ | 2,000 | 16,000 | |
| 18 | $Nb_{56}Mo_{39}Ta_{5}$ | 2,000 | 16,000 | |

| | | | |
|---|---|---|---|
| 19 | $Nb_{73}Mo_{22}Ta_5$ | 2,000 | 16,000 |
| 20 | $Nb_{10}Mo_{80}Ta_{10}$ | 2,000 | 16,000 |
| 21 | $Nb_{27}Mo_{63}Ta_{10}$ | 2,000 | 16,000 |
| 22 | $Nb_{44}Mo_{46}Ta_{10}$ | 2,000 | 16,000 |
| 23 | $Nb_{61}Mo_{29}Ta_{10}$ | 2,000 | 16,000 |
| 24 | $Nb_{15}Mo_{70}Ta_{15}$ | 2,000 | 16,000 |
| 25 | $Nb_{33}Mo_{52}Ta_{15}$ | 2,000 | 16,000 |
| 26 | $Nb_{51}Mo_{34}Ta_{15}$ | 2,000 | 16,000 |
| 27 | $Nb_{20}Mo_{60}Ta_{20}$ | 2,000 | 16,000 |
| 28 | $Nb_{40}Mo_{40}Ta_{20}$ | 2,000 | 16,000 |
| 29 | $Nb_{90}Mo_5Ta_5$ | 2,000 | 16,000 |
| 30 | $Nb_{73}Mo_5Ta_{22}$ | 2,000 | 16,000 |
| 31 | $Nb_{56}Mo_5Ta_{39}$ | 2,000 | 16,000 |
| 32 | $Nb_{39}Mo_5Ta_{56}$ | 2,000 | 16,000 |
| 33 | $Nb_{22}Mo_5Ta_{73}$ | 2,000 | 16,000 |
| 34 | $Nb_{80}Mo_{10}Ta_{10}$ | 2,000 | 16,000 |
| 35 | $Nb_{63}Mo_{10}Ta_{27}$ | 2,000 | 16,000 |
| 36 | $Nb_{46}Mo_{10}Ta_{44}$ | 2,000 | 16,000 |
| 37 | $Nb_{29}Mo_{10}Ta_{61}$ | 2,000 | 16,000 |
| 38 | $Nb_{70}Mo_{15}Ta_{15}$ | 2,000 | 16,000 |
| 39 | $Nb_{52}Mo_{15}Ta_{33}$ | 2,000 | 16,000 |
| 40 | $Nb_{34}Mo_{15}Ta_{51}$ | 2,000 | 16,000 |
| 41 | $Nb_{60}Mo_{20}Ta_{20}$ | 2,000 | 16,000 |
| 42 | $Nb_{40}Mo_{20}Ta_{40}$ | 2,000 | 16,000 |
| 43 | $Nb_{50}Mo_{25}Ta_{25}$ | 2,000 | 16,000 |
| 44 | $Nb_{25}Mo_{50}Ta_{25}$ | 2,000 | 16,000 |
| 45 | $Nb_{25}Mo_{25}Ta_{50}$ | 2,000 | 16,000 |
| 46 | $Nb_{33}Mo_{33}Ta_{33}$ | 2,000 | 16,000 |

## Supplementary References